\documentclass[conference,compsoc]{IEEEtran}
\usepackage{hyperref}

\ifCLASSINFOpdf
\else
\fi
\usepackage{amsmath}
\usepackage{algorithm}
\usepackage{algorithmic}

\usepackage{color}
\usepackage{xcolor}

\definecolor{codegreen}{rgb}{0,0.6,0}
\definecolor{codegray}{rgb}{0.5,0.5,0.5}
\definecolor{codepurple}{rgb}{0.58,0,0.82}
\definecolor{codebackground}{rgb}{0.95,0.95,0.95}

\definecolor{hyperlink-deepgreen}{rgb}{0,100,0}
\usepackage{hyperref}
\hypersetup{
    colorlinks=true,
    linkcolor=blue,
    filecolor=magenta,      
    urlcolor=cyan,
    citecolor=hyperlink-deepgreen,
}
\usepackage{listings}

\usepackage{listings}
\lstdefinestyle{codes}{
    backgroundcolor=\color{codebackground},   
    commentstyle=\color{codegreen},
    keywordstyle=\color{magenta},
    numberstyle=\tiny\color{codegray},
    stringstyle=\color{codepurple},
    basicstyle=\small\ttfamily,
    breakatwhitespace=false,         
    breaklines=true,                 
    keepspaces=true,                 
    numbers=left,                    
    numbersep=5pt,                  
    showspaces=false,                
    showstringspaces=false,
    showtabs=false,                  
    tabsize=2
}
\usepackage{tikz}
\newcommand{\inlinecode}[1]{%
  \tikz[baseline=(X.base)]\node [font=\small\ttfamily, fill=codebackground, text=black, rectangle, rounded corners=1.5pt, inner sep=2pt, text height=2mm, text depth=0.5mm] (X) {#1};%
}

\usepackage[backend=biber, style=numeric, citestyle=numeric, sorting=none]{biblatex}
\usepackage{amsmath}
\usepackage{enumitem} 
\usepackage{amsmath}
\usepackage{multirow}
\usepackage{algorithm}
\usepackage{algorithmic}
\usepackage{makecell}
\usepackage{subcaption}

\renewcommand{\footnoterule}{
    \kern -3pt
    \hrule width 0.5\textwidth height 0.4pt
    \kern 2.6pt
}
\usepackage{cleveref}
\usepackage{float}

\usepackage{makecell}
\usepackage{longtable}
\usepackage{multirow}
\usepackage{caption}
\usepackage{booktabs}

\begin{document}

%
\title{MDHP-Net: Detecting an Emerging Time-exciting Threat in IVN}

\author{Qi Liu$^\dagger$, Yanchen Liu$^\dagger$, Ruifeng Li, Chenhong  Cao, Yufeng Li$^*$, Xingyu Li$^*$, Peng Wang, Runhan Feng, Shiyang Bu}
\maketitle

\begin{abstract}
The integration of intelligent and connected technologies in modern vehicles, while offering enhanced functionalities through Electronic Control Unit (ECU) and interfaces like OBD-II and telematics, also exposes the vehicle’s in-vehicle network (IVN) to potential cyberattacks. 
Unlike prior work, we identify a new time-exciting threat model against IVN. These attacks inject malicious messages that exhibit a time-exciting effect, gradually manipulating network traffic to disrupt vehicle operations and compromise safety-critical functions. 
We systematically analyze the characteristics of the threat: dynamism, time-exciting impact, and low prior knowledge dependency. To validate its practicality, we replicate the attack on a real Advanced Driver Assistance System via Controller Area Network (CAN), exploiting Unified Diagnostic Service vulnerabilities and proposing four attack strategies. While CAN’s integrity checks mitigate attacks, Ethernet migration (e.g., DoIP/SOME/IP) introduces new surfaces. We further investigate the feasibility of time-exciting threat under SOME/IP.
To detect time-exciting threat, we introduce MDHP-Net, leveraging Multi-Dimentional Hawkes Process (MDHP) and temporal and message-wise feature extracting structures. Meanwhile, to estimate MDHP parameters, we developed the first GPU-optimized gradient descent solver for MDHP (MDHP-GDS). These modules significantly improves the detection rate under time-exciting attacks in multi-ECU IVN system.
To address data scarcity, we release STEIA9, the first open-source dataset for time-exciting attacks, covering 9 Ethernet-based attack scenarios. Extensive experiments on STEIA9 (9 attack scenarios) show MDHP-Net outperforms 3 baselines, confirming attack feasibility and detection efficacy.

\end{abstract}


%
\IEEEpeerreviewmaketitle

\setlength{\parindent}{1em}

\section{Introduction}
Connected and automated vehicles (CAVs) integrate numerous Electronic Control Units (ECUs), which are interconnected through various types of In-Vehicle Networks (IVNs), such as Controller Area Networks (CAN), Local Interconnect Networks, and FlexRay, to control various vehicle functionalities. Furthermore, interfaces including the OBD-II port, sensors, and telematics facilitate communication between the IVN and the external environment. These interfaces provide notable benefits, such as mitigating traffic congestion, enhancing driving convenience, and promoting driving safety \cite{pereira2017automated}.

Attacks on the IVN or ECUs pose significant threats to the safety and stability of CAVs. Many investigations have focused on identifying and analyzing the vulnerabilities of CAVs \cite{chen2023towards}\cite{sun2021survey}\cite{de2024systematic}. As CAVs are increasingly equipped with various interfaces (e.g., OBD-II, TCU, Gateway) to interact with external entities (e.g., diagnostic tools, remote servers, and telematics systems), attackers are finding new ways to exploit these interfaces and attack the vehicles. For instance, vulnerabilities in the OBD-II interface allow adversaries to inject malicious commands into the low-speed and high-speed CAN buses, affecting critical systems such as body control, powertrain, and safety systems \cite{xue2022said}\cite{clifford2016}. Additionally, researchers have shown that vulnerabilities in aftermarket vehicle diagnostic tools can be exploited to remotely control various vehicle functions, such as doors, windows, and mirrors \cite{lyu2016}. Notable examples include security researchers demonstrating vulnerabilities in Jeep \cite{miller2015remote}, BMW \cite{keenlab2018} and Tesla's Model S and X \cite{keenlab2019}, successfully hacking into vehicle systems to remotely control various functions, including doors, brakes, and keyless entry. Each of these attacks was executed by injecting specific data into the vehicle’s IVN through exposed interfaces.

Distinct from prior works\cite{yu2022towards, ruan2025picacan, huang2024finebid, lin2024phade, wang2024distributed, sato2025realism, hu2024l_, wang2024resilient, liu2024distributed, jia2024magic, wang2023optimal, xue2022said}, we identify a time-exciting threat model in IVN. Insights from experienced hackers suggest that attackers, by persistently probing with various tools, such as injection attacks, are more likely to identify vulnerabilities and exploit the system’s weakness. In this scenario, the likelihood of a successful attack is influenced by previous attempts, as past failures accumulate, thereby increasing the probability of success in subsequent attempts. We validate the attack’s feasibility by replicating it on a real Advanced Driver Assistance System (ADAS) via the CAN bus, exploiting Unified Diagnostic Service (UDS) vulnerabilities. 

To identify injection attacks in IVN, existing countermeasures can be divided into two categories \cite{CPY2024research}: the security enhancement of IVN protocols and intrusion detection. The security enhancement of IVN protocols often uses cryptographic approaches to prevent unauthorized access and IVN data tampering \cite{woo2014practical} \cite{lee2023protecting} \cite{lim2016efficient} \cite{wei2016authenticated} \cite{iorio2020securing}\cite{carvajal2021semi}.  However, it suffers from the following shortcomings. First, vehicle operations are subject to strict real-time constraints, making high latency and overhead difficult to tolerate. Second, the incompatibility of security enhancement schemes for in-vehicle network protocols requires manufacturers to redesign ECU firmware, incurring significant costs for suppliers when releasing new firmware. 

Intrusion detection employs a variety of features, such as the entropy of in-vehicle network\cite{muter2011entropy}, message intervals \cite{miller2013adventures}, message correlation/consistency \cite{muter2010structured}, ECU fingerprints \cite{cho2016fingerprinting}, vehicle voltage \cite{xun2021vehicleeids}, communication characteristics \cite{choi2018voltageids}\cite{alkhatib2021some}\cite{alkhatib2023here}etc., serving as crucial indicators in identifying deviations from normal behavior within the vehicle network. Nevertheless, existing intrusion detection approaches are deployed in IVN and detect anomalies with the stability features of IVN traffic, they cannot identify the injected abnormal messages with normal traffic patterns, especially for time-exciting threat. To overcome the limitation of existing works, in this paper, we propose MDHP-Net to detect time-excitating threat in IVN. MDHP-Net uses Multi-Dimensional Hawkes Process (MDHP) to profile the abnormal behaviors of attacker, and temporal structure MDHP-LSTM is combined to extract time-excitation features. 

Validating the feasibility of time-exciting threat model presents non-trivial challenges.  \textbf{C1}: The first challenge involves empirical verification of this threat model in real-world environments. Reproducing realistic attack scenarios requires a comprehensive analysis to delineate the full scope of the model’s capabilities, attack vectors, and security implications—an area overlooked in prior work. \textbf{C2}: The second challenge lies in detecting such threat models, which entails two major difficulties. Firstly, deriving accurate closed-form solutions for the MDHP remains a challenge, as does the lack of open-source solvers for rapid parameter estimation. Closed-form solutions are essential for precise modeling of the MDHP, enabling a better representation of the dynamic interactions in sequential events. Open-source solvers are equally crucial, as they facilitate quick parameter estimation, which is vital for real-time applications. Secondly, effectively incorporating the feature of attackers' behavior captured by the MDHP into a temporal deep-learning structure remains challenging. Current approaches struggle to seamlessly integrate the self-exciting, event-dependent features of the Hawkes process within the LSTM framework, limiting the model’s ability to enhance temporal feature extraction. Addressing this gap is crucial for improving the analysis and prediction of complex attack patterns over time.

We first address \textbf{C1} by employing a dual-methodological approach: First, we conduct a systematic literature review coupled with theoretical derivation to comprehensively analyze the security implications. Subsequently, we perform empirical validation through CAN bus attacks on operational Advanced Driver Assistance System (ADAS) , where we design four distinct attack vectors exploiting vulnerabilities in UDS, thereby conclusively demonstrating the practical viability and critical risks of this threat model. While CAN's integrity checks offer partial attack mitigation, migration to Ethernet protocols  expands attack surfaces due to their complex, service-oriented architecture. We  further investigate the feasibility of time-exciting threat under Ethernet SOME/IP. We then leverage novel mathematic modeling and designs to address \textbf{C2}: we leverage novel mathematic modeling and designs to address the second challenge. Specifically, we derive the closed-form of the MDHP function to characterize the dynamic behavior of attackers, and further develop a gradient descent solver (MDHP-GDS) tailed for MDHP. A set of advanced optimization strategies is used to obtain the optimal MDHP parameters. Besides, we propose an injection attack detector that integrate estimated MDHP parameters with a novel structure, MDHP-LSTM, to enhance temporal feature extraction. We further adopt a residual self-attention mechanism to capture the relationship between different messages and consequentially accelerate the model’s convergence and enhance detection accuracy.

We summarize the main contribution as follows:
\begin{itemize}[topsep=0pt, left=0pt]
\item We systematically analyze the threat to understand its features. To validate its practicality, we replicate the attack on a real ADAS system via CAN bus, exploiting UDS service vulnerabilities, and proposing four attack strategies.
\item We extend our investigation to SOME/IP, analyzing time-exciting attack feasibility. To detect such attacks, we propose MDHP-Net, a temporal deep learning model combining: (1) MDHP-GDS, an optimized gradient descent solver for efficient MDHP parameter estimation; (2) MDHP-LSTM blocks with residual self-attention to extract attack features and improve model performance. 
The code can be accessed at: \url{https://github.com/Tiara8735/MDHP-Net-Anonymous}.
\item We release STEIA9, the first open-source dataset for time-exciting attacks, covering 9 Ethernet-based attack scenarios. Experimental results indicate that MDHP-GDS achieves high estimation speeds and clearly distinguishes between injection attack scenario distributions. Additionally, MDHP-Net demonstrates robust detection capabilities against time-exciting attacks on IVN, outperforming baseline methods.
\end{itemize}

\section{Background}
\subsection{SOME/IP Protocol}
\begin{figure}[H]
    \centering
    \includegraphics[width=0.95\linewidth]{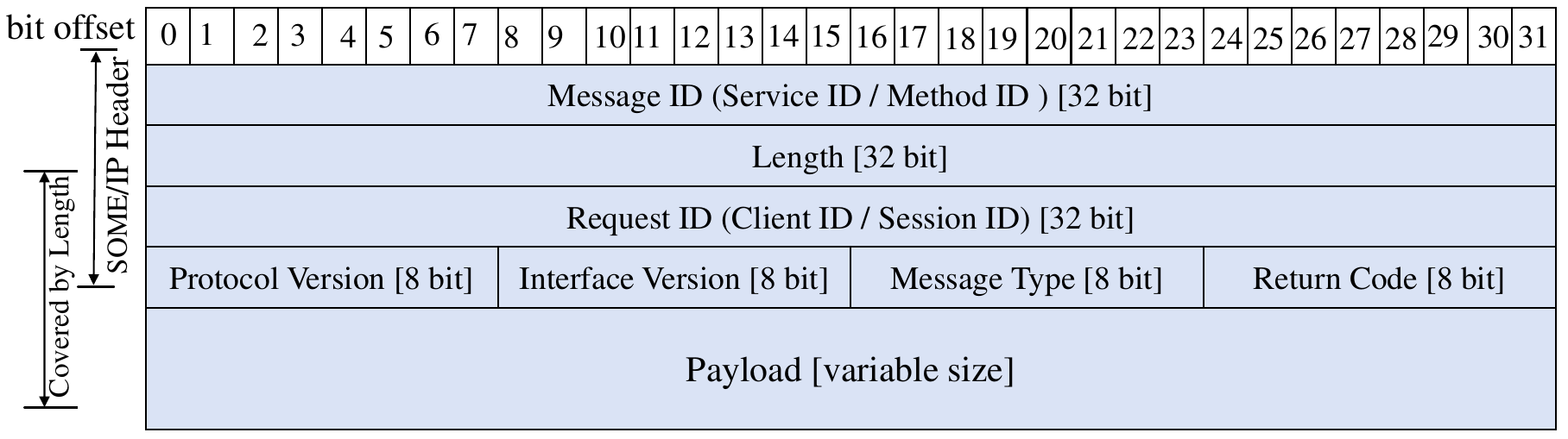}
   \caption{SOME/IP Message Format.}
     \label{fig:SOME/IP protocol}
\end{figure}

SOME/IP (Scalable service-Oriented Middleware over IP) is an IP-based, service-oriented protocol for Ethernet, contrasting with legacy in-vehicle protocols like CAN/MOST. It provides service-oriented features (e.g., event notifications, RPC) and defines its serialization/wire format. Figure \ref{fig:SOME/IP protocol} shows its message format. 

SOME/IP supports three communication models: Request/Response, Fire \& Forget, and Notification Events\cite{AUTOSAR-Protocol}.\\
\textbf{Request/Response}: This communication pattern follows the RPC model, consisting of a request and a corresponding response. \\ 
\textbf{Fire \& Forget}: A type of request where the client sends a message without expecting any response from the server.  \\
\textbf{Notification Events}: Servers push event messages to subscribed clients without explicit client requests.\\

\subsection{Multi-Dimensional Hawkes Process}
Hawkes A.G. \cite{hawkes1971point}\cite{hawkes1971spectra} introduced the concept of Hawkes process and Multi-Dimensional Hawkes process in 1971. The essential property of the Hawkes process is that the occurrence of any event increases the probability of subsequent events occurring. Here, we present some definitions related to Hawkes
Process.\\
\textbf{\textit{Definition 1:}} A counting process is a specific type of stochastic process\(\{N(t), t \geq 0\}\) that satisfies the condition where \(N(t)\) denotes the cumulative number of events that have occurred by time \(t\) \cite{ross1995stochastic} \cite{dutta2020hawkeseye}. \\
\textbf{\textit{Definition 2:}} A point process is a type of stochastic process consisting of sparse events occurring in time or space \cite{mcfadden1965entropy}. Its sparsity is reflected in the fact that, within any sufficiently small time interval, at most one event can occur.\\
\textbf{\textit{Definition 3:}} The intensity function \(\lambda(t)\) , describes the instantaneous rate at which events occur within a point process.\\
\textbf{\textit{Definition 4:}} The conditional intensity function is more generalized as it takes into account the historical information of the process\footnote{Some papers exhibit inconsistencies in their use of the terms "intensity function" and "conditional intensity function."  In many cases, the term "intensity function" is used for simplicity, and readers are expected to infer from the context that it refers to a conditional intensity function.}.\\
\textbf{\textit{Definition 5:}} Point processes that possess an intensity function and also serve as counting processes are known as regular point processes(RPPs) \cite{rubin1972regular}. 

The Hawkes process is a prominent example of an RPP, with its intensity function defined as follows \cite{hawkes1971spectra}:
\begin{align}\label{eq:intensity function-Hawkes process}
    \lambda(t) = \mu + \int_0^t g(t-s)\, dN(s) \notag &= \mu + \sum_{x^{\tau} < t} g(t - x^{\tau})  \\
               &= \mu + \sum_{x^{\tau} < t} \alpha e^{-\beta(t - x^{\tau})} 
\end{align}

\setlength{\parindent}{0em}
where \(\mu\) is the the baseline intensity of the Hawkes process; \(g(t)\) is the excitation function which is typically defined as \(\alpha e^{-\beta(t)}\) \cite{ozaki1979maximum}.
\setlength{\parindent}{1em}



\begin{figure}[!htbp]
    \centering
  \includegraphics[width=0.8\linewidth]{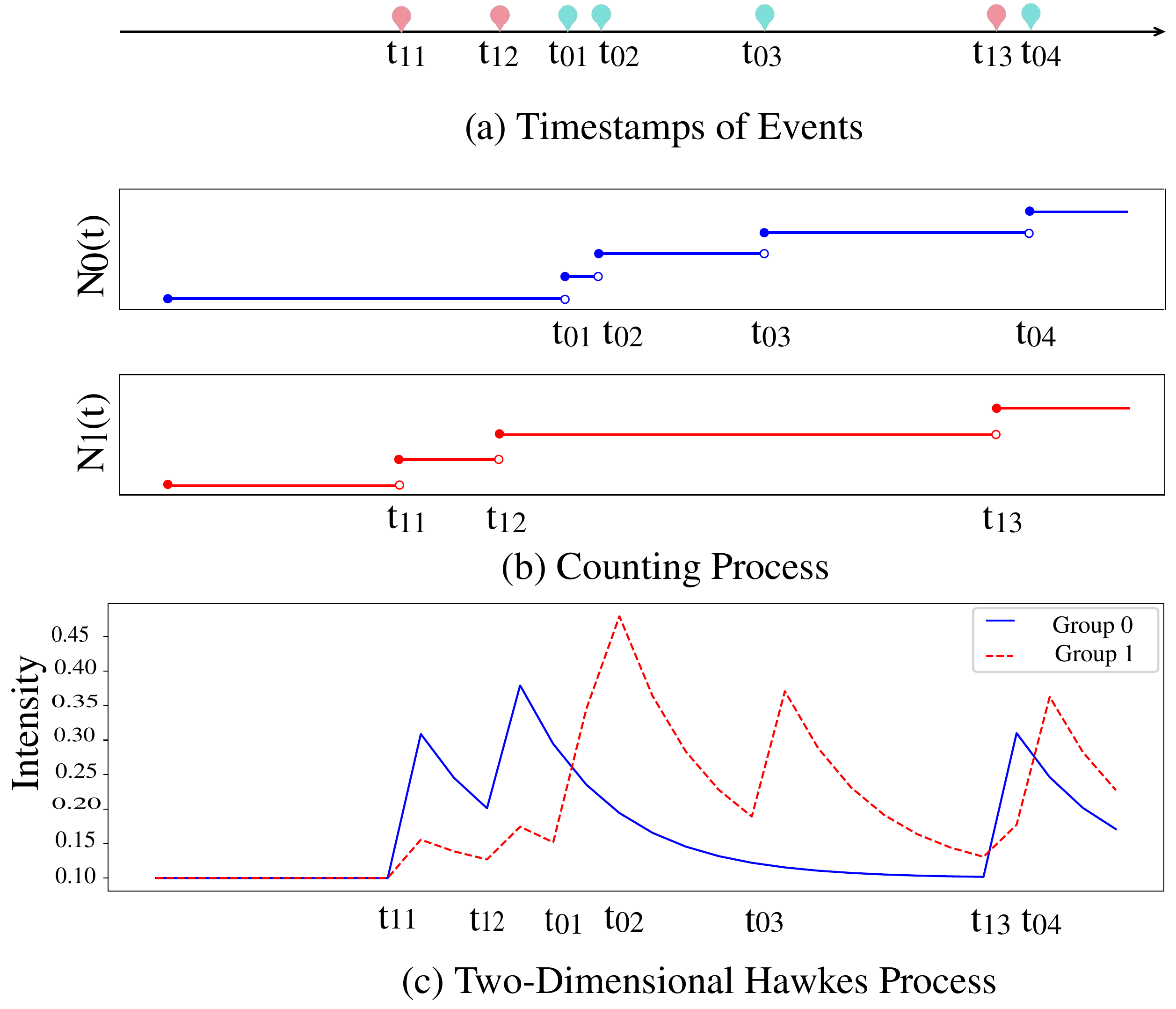}
    \caption{Two-Dimensional Hawkes Process: (a) The occurrence of two types of events within a time interval. (b) The counting process of each event type over time. (c) The intensity function of a bivariate Hawkes process.}
    \label{hawkes process-intensity}
\end{figure}

For Multi-Dimensional Hawkes process, we not only consider the self-excitation effect of a single type of event but also the mutually-excitation between different types of events (Figure \ref{hawkes process-intensity}). Specifically, the intensity function can be expressed as:
\begin{equation}\label{eq:intensity function-Multi-Dimensional Hawkes process}
\lambda^{i}(t) = \theta_i + \sum^{D-1}_{j=0}\sum_{k\land T^{k}_{j} < t} \alpha^{(i,j)} e^{-\beta^{(i,j)}(t-T^{k}_{j})}
\end{equation}
where 
\begin{itemize}[topsep=0pt, left=0pt]
    \item \( \theta_i \) represents the baseline intensity of the Mutli-Dimensional Hawkes process, where \( i \) denotes the number of events (equivalent to the number of ECUs). 
    \item \( \alpha^{(i,j)} \) represents the amplitude coefficient of the Mutli-Dimensional Hawkes process, indicating the positive influence of event \( i \)'s occurrence on event \( j \).
    \item \( \beta^{(i,j)} \) represents the decay coefficient of the Mutli-Dimensional Hawkes process.  
\end{itemize}

\section{Time-exciting Threat Model}\label{sec:threat-model}
In this paper, we investigate a specific type of time-exciting threat. This type of threat involves injecting abnormal data or messages into the IVN through external entities such as telematics systems, OBD-II ports, T-Boxes, and various sensors. The primary objective of this threat model is to exploit these entry points to compromise the security of IVN systems and achieve specific malicious targets.

The attacker possesses certain access resources and permissions, allowing them to make repeated attempts to identify vulnerabilities within the system. The likelihood of a successful attack is no longer completely independent of previous attempts. Instead, the experience gained from past attempts enhances the chances of a successful attack over time. This reflects a time-exciting effect, where each attempt affect the probability of success in future attacks. 

\subsection{Main features}
In this scenario, the attacker is assumed to possess only a basic understanding of specific messages and data fields within the IVN, including protocols such as Ethernet and CAN, which are commonly employed in automotive systems. The attacker may gain physical access to the IVN via the OBD port or establish a remote connection through an On-Board Unit (OBU). With this level of access, they can monitor and transmit packets across the IVN, enabling them to log, analyze, and inject individual or sequential messages. Such intrusions may disrupt normal vehicle operations, potentially causing unintended malfunctions or inducing the vehicle to execute actions inconsistent with its current state.

The key characteristics of this attack are summarized as follows:
\begin{itemize}[topsep=0pt, left=0pt]
    \item \textbf{Dynamic Nature:} The success probability of the attack varies dynamically over time, with each attempt influencing the likelihood of subsequent successes.
    \item \textbf{Time-exciting Effect:} The time-exciting effect enables the attacker to improve the attack's success rate through repeated attempts, refining the strategy iteratively with experience.
    \item \textbf{Limited Knowledge Requirement:} The attacker does not require in-depth knowledge of the target system’s internal mechanisms; rather, they only need to construct data packets that comply with basic protocol specifications.
\end{itemize}

Based on above characteristics, we compare Time-exciting threat with existing attacks, as summarized in Table \ref{tab:attack_features_comparison}. Existing approaches exhibit distinct operational profiles: reverse engineering and spoofing attacks primarily rely on static analysis and predefined patterns, while DoS attacks operate in a stateless manner. State-dependent injection attacks and APTs demonstrate dynamic capabilities through state awareness or multi-stage execution, though with varying knowledge requirements.The Time-exciting threat model distinguishes itself by integrating temporal adaptability (dynamically evolving success rates), cumulative learning (iterative optimization through experience), and minimal prior knowledge dependencies. Unlike existing methods, which operate under fixed or staged strategies, this model continuously adjusts based on real-time conditions, enabling sustained and evolving attack efficacy. This dynamic nature positions it as a distinct class of threats requiring novel detection and mitigation approaches.

\begin{table*}[htbp]
\centering
\caption{How time-exciting threat model compares with other attacks}
\renewcommand{\arraystretch}{1.0}
\setlength{\tabcolsep}{5pt}
\begin{tabular}{p{6cm}p{2.6cm}p{3.5cm}p{2.5cm}}
\hline
\textbf{Attack Type} & \textbf{Dynamic} & \textbf{Time-exciting Effect} & \textbf{Prior Knowledge} \\
\hline
Time-Exciting Threat Model & 
High & 
Yes & 
No \\
Reverse Engineering Attacks\cite{yu2022towards}\cite{ruan2025picacan}\cite{huang2024finebid} & 
Low & 
Low & 
Yes \\
Spoofing Attacks\cite{lin2024phade}\cite{wang2024distributed}\cite{sato2025realism} & 
Low & 
Low & 
Partial \\
DoS Attacks\cite{hu2024l_}\cite{wang2024resilient}\cite{liu2024distributed} & 
Low  & 
None & 
No \\
State-aware Injection Attacks\cite{xue2022said} & 
High & 
Partial & 
Yes\\
Advanced Persistent Threats\cite{jia2024magic}\cite{wang2023optimal} & 
High& 
Yes & 
No \\
\hline
\end{tabular}
\label{tab:attack_features_comparison}
\end{table*}

\subsection{Practical Deployment}

\begin{figure}[!htbp]
    \centering
    \includegraphics[width=0.45\textwidth]{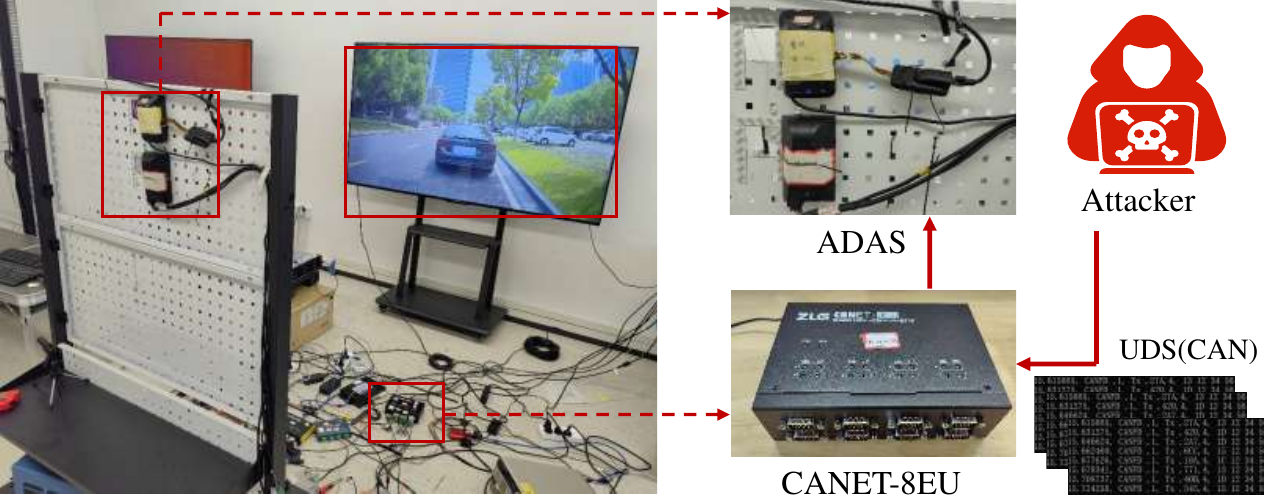}
    \caption{Practical test environment deployment.}
    \label{fig:deploy}
\end{figure}

\textbf{Attack Vector Initialization}. The attack was initiated by injecting fabricated road scene videos into the ADAS camera to simulate real-world driving conditions, while custom CAN messages were used to manipulate ADAS functionality. The testing employed the ZLG/CANET-8E-U high-performance Ethernet-to-CAN module, featuring 8 CAN channels compliant with CAN 2.0A/B standards, with a transmission capacity of 8,000 frames/sec per channel and a reception capacity of 12,000 frames/sec. The module supports 10/100/1000 Mbps auto-negotiated Ethernet with a forwarding latency of <5 ms.\\
\textbf{Attack Execution}. Figure \ref{fig:deploy} illustrates how malicious actors can compromise ADAS systems through either physical access (OBD-II interface) or remote network infiltration, targeting critical CAN bus communications and message routing mechanisms. The implemented attack vectors consist of (1) network fuzzing attacks employing exhaustive and stochastic randomization of CAN message identifiers and payload data, and (2) UDS service attacks involving the malicious generation and injection of ISO 14229-compliant diagnostic packets. These sophisticated attack methodologies ultimately lead to critical system failures including sensor perception anomalies, decision-making logic corruption, and complete functional degradation of ADAS capabilities.\\
\begin{figure}[htbp]
  \centering
  \begin{subfigure}[t]{0.4\textwidth}
    \centering
    \includegraphics[width=\linewidth]{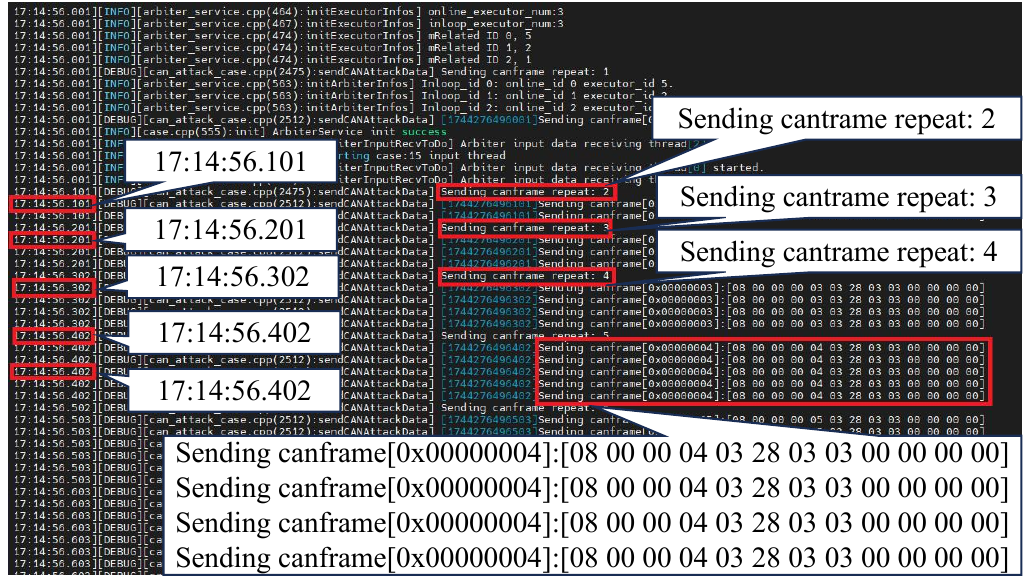}
    \caption{Attack message data visualization}
    \label{fig:Attack message data visualization}
  \end{subfigure}
  \hfill
  \begin{subfigure}[t]{0.4\textwidth}
    \centering
    \includegraphics[width=\linewidth]{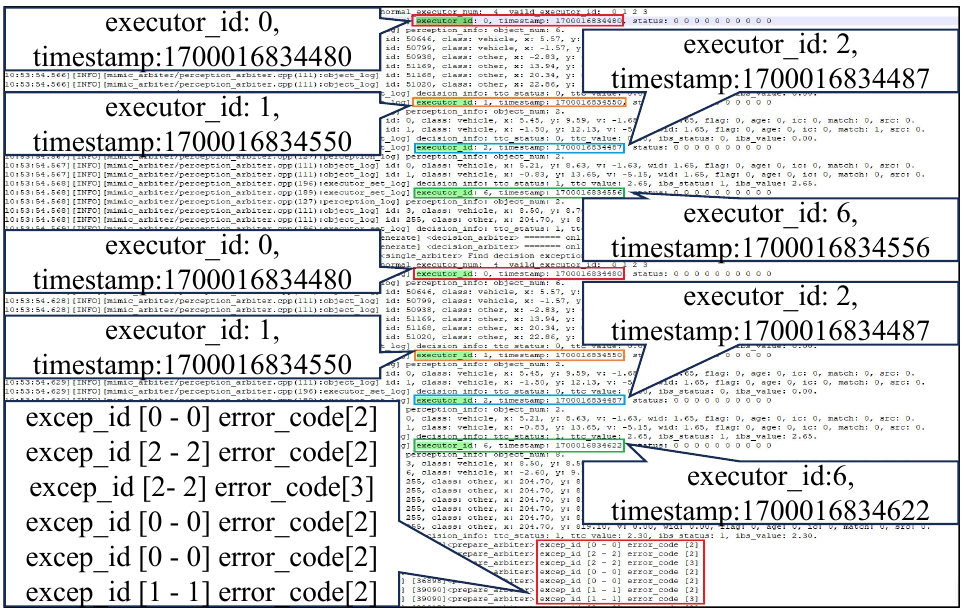}
    \caption{Impact Analysis of the Attack}
    \label{fig:Attack effect}
  \end{subfigure}
  \caption{Verification of Attack Impact.}
  \label{fig:combined}
\end{figure}
\textbf{Impact Assessment}. Each CAN attack was logged, recording the attack payload and frame count per attack cycle. As highlighted in Figure \ref{fig:Attack message data visualization} (red box), the attack interval was set to 100 ms, with an incremental escalation in attack intensity (i.e., +1 CAN frame per iteration). 
Figure \ref{fig:Attack effect} demonstrates the anomalous effects induced on targeted ADAS components. Specifically, the execution units with ID 0.1.2 (the exclusive attack targets) exhibit timestamp freezing at a fixed value, accompanied by complete cessation of data output. In contrast, non-targeted ADAS units (e.g., ID 6) maintain normal operation with real-time timestamp updates. The diagnostic message "excep\_id [0 - 0] error\_code[2]" indicates an output data anomaly (Type 2 error) in the first executor (ID 0), thereby experimentally validating the feasibility of time-excitation attacks.

A detailed exposition of these advanced attack methodologies will be presented in the subsequent section.

\subsection{Time-exciting attack strategies}

For time-exciting threat, we design four attack strategies as shown in Table\ref{tab:Attack strategy}:\\
\textbf{PLA - Power-law acceleration strategy}: Unlike constant-rate injection attacks\cite{liu2024sissa}\cite{8937816}\cite{seo2018gids}\cite{8688625}(no time-exciting), this strategy adopts a power-law-based injection mechanism, wherein the attack rate evolves over time according to a power-law distribution. Initially, the injection rate remains low, closely resembling legitimate traffic, which makes early detection significantly more difficult. As time progresses, the attacker gradually amplifies the attack intensity.  Here, \( a \) represents the overall attack intensity, determining the amplitude of the injection rate, while \( b \) denotes the power-law exponent, which shapes the growth trend of the rate.   \\
\textbf{DEA - Delayed escalation attack strategy}: In contrast to the Power-law acceleration strategy, this strategy more closely reflects the behavior patterns of real-world attackers. The attacker begins with a low injection rate, performing a slow attack at the outset. Once the attacker gathers more information and confirms that the attack remains undetected (at time $t_1$), they quickly release resources and escalate the injection exponentially to achieve their attack objective. In $g(t)$, \( W_1 \) and \( W_2 \) are scaling factors for the two distinct phases of the attack. \( \alpha_1 \) determines the intensity of the power-law increase during the initial phase, and \( \alpha_2 \) governs the rate of exponential growth in the later stage. \\
\textbf{ASA - Adaptive stealth-oriented attack strategy}: This strategy takes into account not only the attacker’s resource accumulation but also the defense mechanisms of the target. Initially, the attacker injects packets at a steady rate, gradually increasing until reaching a peak. Before and after the peak, we consider the dynamic response of the defense system (e.g., traffic filtering\cite{argyraki2005active}), which forces the attacker to reduce the attack frequency, leading to a decrease in the attacker’s rate of injection. In $g(t)$, \( C \) acts as an overall scaling factor to adjust the peak height. \( \gamma \) controls the growth speed, and \( t_0 \) determines the timing of the peak attack intensity. \\
\textbf{DAM - Dynamic adjustment via multi-strategy integration}: We propose an attack strategy that integrates above-mentioned strategies and can dynamically adjust its parameter $w$. In this strategy, $\alpha_1$ and $\alpha_2$ are both greater than 1, where \( \alpha_1 \) controls the growth of the power-law component, while \( \alpha_2 \) influences the intensity of the exponential surge. \( w \in [0,1] \) is used to control the weight.  For example, given specific values for $\alpha_1$ and $\alpha_2$, we assume that $\alpha_1$ corresponds to a low-rate attack and $\alpha_2$ corresponds to a high-rate attack. As $w$ decreases, the attack strategy gradually shifts towards a high-rate attack; conversely, as $w$ increases, the strategy leans toward a low-rate attack. By adjusting the weighting parameter, the attacker is able to flexibly switch among different attack strategies, including probing, suppression, and hybrid patterns.
\begin{table}[!htbp]
\centering
\caption{Different Time-exciting injection attack strategies and their corresponding attack rate forms} 
\setlength{\tabcolsep}{3pt}
\begin{tabular}{cc}
\hline
\textbf{Attack strategy} & \textbf{Attack rate }  \\ \hline
 Power-law acceleration & $g(t)=a\cdot t^b$ \\ \hline
 \multirow{3}{*}{\centering\shortstack{Delayed escalation\\attack}} & 
\multirow{2}{*}{$g(t)=
\begin{cases}
W_1\cdot \alpha_1\, t^{\alpha_1-1}, & t < t_1, \\[1mm]
W_2\cdot \alpha_2\, e^{\gamma(t-t_1)}, & t \ge t_1.
\end{cases}$} \\
 & \\& \\ \hline
 \multirow{2}{*}{\shortstack{Adaptive stealth-oriented \\attack}}   & 
\multirow{2}{*}{$g(t)=C\cdot \frac{e^{\gamma t}}{\left(1+e^{\gamma (t-t_0)}\right)^2}$}\\ & \\ \hline
\multirow{2}{*}{\shortstack{Dynamic adjustment via\\multi-strategy integration}} & 
\multirow{2}{*}{$g(t)=w\cdot \alpha_1\,t^{\alpha_1-1}+\left(1-w\right)\cdot \alpha_2\, e^{\alpha_2t}$} \\
& \\ \hline
\end{tabular}
\label{tab:Attack strategy}
\end{table}

\section{The Proposed Detection Model}
\label{MDHP-NET}
\subsection{Overview}
\begin{figure}[htbp]
    \centering
\includegraphics[width=0.5\textwidth]{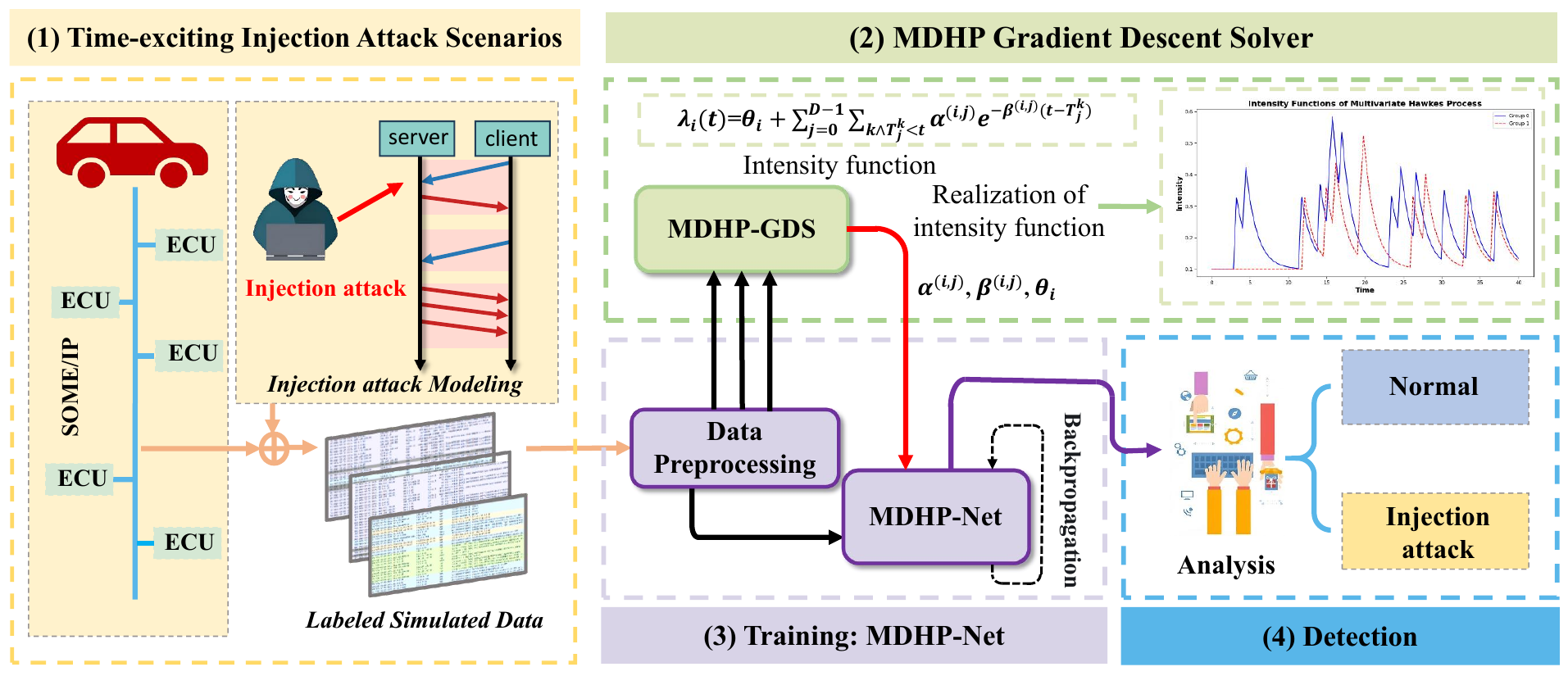}
    \caption{The detection framework for time-exciting threat.}
    \label{fig:detection framework}
\end{figure}

We propose a detection model (Fig. \ref{fig:detection framework}) that profiles IVN anomalies, using SOME/IP as our attack scenario case study. While SOME/IP overcomes traditional automotive protocols' (CAN) bandwidth limitations, vulnerabilities in commercial stacks (Bluetooth, Wi-Fi, and 4G modules) create unauthorized IVN access risks.

To capture the dynamic behavior of time-exciting injection attacks, we first outline these attack scenarios and provide a precise derivation of closed-form solutions. We then develop an open-source gradient descent solver tailored for MDHP to enable rapid parameter estimation. Finally, we integrate optimal MDHP parameters into MDHP-Net, incorporating key attacker behavior characteristics, which allows us to classify and detect potential attacks accurately.

\subsection{MDHP Gradient Descent Solver}
\subsubsection{\textbf{Likelihood}}\label{mdhp-gds-likelihood}

The log-likelihood function for the univariate Hawkes process has been given by T.Ozaki.\cite{ozaki1979maximum}. For Multi-Dimensional Hawkes processes, we obtained the most comprehensive log-likelihood function through the detailed derivation below. Initially, the likelihood function for Multi-Dimensional Hawkes processes is expressed by a straightforward equation: 


\begin{equation}\label{eq:L}
 L = \prod_{i=0}^{D-1} \prod_{t=T_i^{(0)}}^{T_i^{\text{last}}} {\lambda}^i(t) \cdot e^{-\int_0^{T_{\text{span}}} {\lambda}^i(v) dv}
\end{equation}

\setlength{\parindent}{0em}
where \({\lambda}^i(t) \) is the intensity function of each ECU. The full derivation is given in Appendix \ref{Appdenix A}.The likelihood function is further transformed into the log-likelihood:
\setlength{\parindent}{1em}

\begin{equation}\label{eq:log-likelihood}
	\begin{aligned}
		\ln L &= \sum_{i=0}^{D-1} \sum_{t=T_i^{(0)}}^{T_i^{\text{last}}} \ln \left( \theta_i+\sum_{j=0}^{D-1} \sum_{k : T_j^k < t} \alpha^{(i,j)} e^{-\beta^{(i,j)} (t-T_j^k ) } \right) \\ &\quad -\underbrace{\sum_{i=0}^{D-1} \int_{0}^{T_{\text{span}}} \left( \theta_i+\sum_{j=0}^{D-1} \sum_{k : T_j^k < v} \alpha^{(i,j)} e^{-\beta^{(i,j)} (v-T_j^k ) } \right) dv}_{\text{call this } \Gamma} 
	\end{aligned}
\end{equation}

In fact, \(\Gamma\) is a computable function. Therefore, we further simplify the expression to obtain a more accurate log-likelihood function, as shown in Equation \ref{eq:mdhp-gds-ln-likelihood}. The full derivation is given in Appendix \ref{Appdenix B}.

\setlength{\abovedisplayskip}{0pt}
\begin{equation}\label{eq:mdhp-gds-ln-likelihood}
\begin{aligned}
\ln{L}&=\sum_{i=0}^{D-1}\sum_{t=T_i^{(0)}}^{T_i^\text{last}}\ln\left(\theta_i+\sum_{j=0}^{D-1}\sum_{k \land T_j^k<t}\alpha^{(i,j)}e^{-\beta^{(i,j)}\left(t-T_j^k\right)}\right) \\ &
-T_\text{span}\sum_{i=0}^{D-1}\theta_i \\&
+ \sum_{i=0}^{D-1}\sum_{j=0}^{D-1}\left(\frac{\alpha^{(i,j)}}{\beta^{(i,j)}}\sum_{k}\left(e^{-\beta^{(i,j)}\left(T_\text{span}-T_j^{(k)}\right)}-1\right)\right)
\end{aligned}
\end{equation}

\setlength{\parindent}{0em}
where
\setlength{\parindent}{1em}

$$
\begin{aligned}
D =& ~ \text{MDHP Dim, i.e., Number of ECUs} \\
T_i^{(k)} =& ~ \text{Timestamp of the $k^\text{th}$ message sent by the $i^\text{th}$ ECU} \\
T_\text{span} =& ~ \text{Time span of the observation window} \\
\alpha, \beta, \theta =
& ~ \text{Hawkes parameters;} \\
& \,\, \text{Both \(\alpha\) and \(\beta\) are \((D, D)\) matrices,} \\
& \,\, \text{and \(\theta\) is a \((D)\) vector.}
\end{aligned}
$$


\subsubsection{\textbf{The Design of MDHP-GDS}}

We develop a gradient descent solver, MDHP-GDS\footnote{We provide a comprehensive Jupyter notebook at the following link: \url{https://github.com/Tiara8735/MDHP-Net-Anonymous/blob/main/examples/MDHP-GDS.ipynb}}, for estimating the parameters of Multi-Dimensional Hawkes Process.

MDHP-GDS is implemented entirely in Python and utilizes PyTorch's Autograd feature extensively \cite{pytorch-auto-grad}. For our specific application, the parameters to be updated are $\alpha$, $\beta$ and $\theta$, using the negative log-likelihood ($-\ln{L}$) as the loss function in Equation \ref{eq:mdhp-gds-ln-likelihood}.

\begin{figure}[!htbp]
    \centering
    \includegraphics[width=0.45\textwidth]{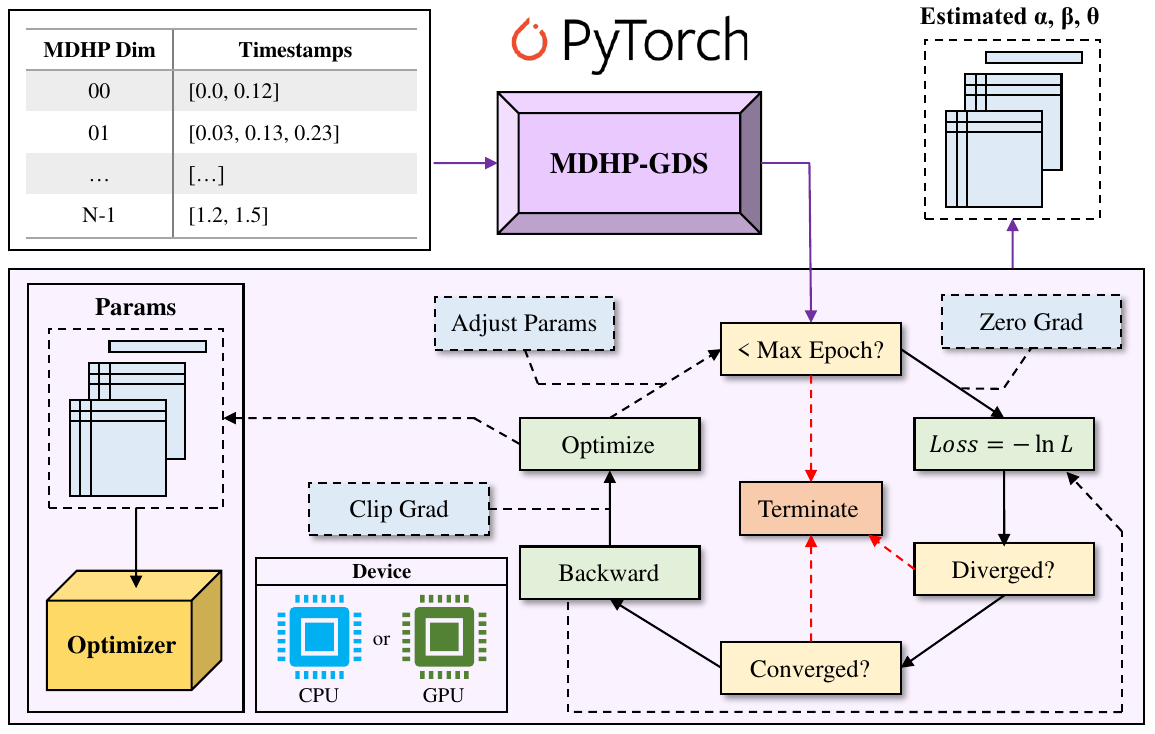}
    \caption{Workflow of MDHP-GDS.}
    \label{fig:mdhp-gds-workflow}
\end{figure}

Workflow of MDHP-GDS is depicted in Figure \ref{fig:mdhp-gds-workflow}. The solver contains three tensors, two of shape \inlinecode{(dim, dim)}, and one of shape \inlinecode{(dim)}, representing the parameters $\alpha$, $\beta$ and $\theta$. These tensors are Autograd-enabled, allowing them to be learnable parameters that can be optimized using a PyTorch optimizer.

Calculating the loss function constitutes the most time-consuming part of the gradient descent procedure, and the complex loop structure of Equation \ref{eq:mdhp-gds-ln-likelihood} presents a more severe challenge for fast parameter-estimation speed. Therefore, in the context of in-vehicle communications, where vast quantities of data are transmitted each millisecond, it is crucial to implement a well-conceived optimization strategy for loss calculation in each epoch. 

\subsubsection{\textbf{Optimization Strategies of MDHP-GDS}}\label{mdhp-gds-optimization}

In general, MDHP-GDS incorporates advanced optimization techniques, including: 1. \textit{Padding and Masking}; 2. \textit{Computational Decoupling}; 3. \textit{Vectorization}; 4. \textit{GPU Offloading}. 5. \textit{Just-In-Time (JIT) Compilation}. These methods enhance computational efficiency, and the correlated algorithms are detailed in this section. 

First of all, Equation \ref{eq:mdhp-gds-ln-likelihood} is split into Part1, Part2, and Part3, as shown in the following parts:

$$
\begin{aligned}
\text{Part1} =& \sum_{i=0}^{D-1}\sum_{t=T_i^{(0)}}^{T_i^\text{last}}\ln\left(\theta_i+\sum_{j=0}^{D-1}\sum_{k \land T_j^k<t}\alpha^{(i,j)}e^{-\beta^{(i,j)}\left(t-T_j^k\right)}\right) \\
\text{Part2} =& -T_\text{span}\sum_{i=0}^{D-1}\theta_i \\
\text{Part3} =& \sum_{i=0}^{D-1}\sum_{j=0}^{D-1}\left(\frac{\alpha^{(i,j)}}{\beta^{(i,j)}}\sum_{k}\left(e^{-\beta^{(i,j)}\left(T_\text{span}-T_j^{(k)}\right)}-1\right)\right) \\
\end{aligned}
$$

And $\ln{L}$ can now be described as:

$$
\ln{L} = \text{Part1} + \text{Part2} + \text{Part3}
$$

For each part, as well as the initialization and execution stages, the adopted optimization strategies will be explained in detail in the following sections.

\noindent\textbf{+ Stage1: Initialization}

Initialization of MDHP-GDS begins by receiving a list of tensors as input data. Each tensor corresponds to a Hawkes dimension, and therefore there are \inlinecode{dim} tensors in total.

\begin{figure}[!htbp]
    \centering
    \includegraphics[width=0.5\textwidth]{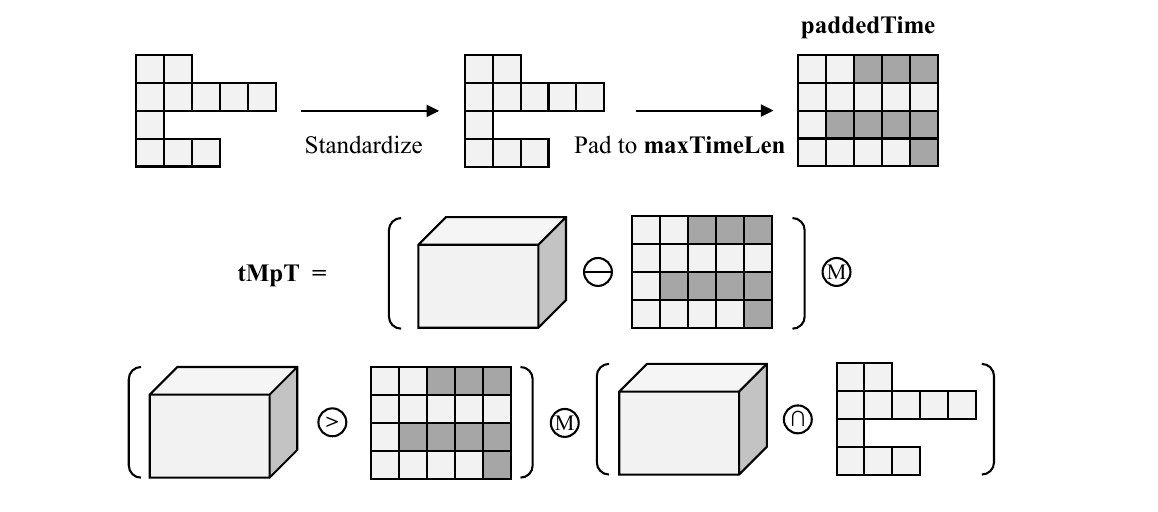}
    \caption{Initialization process.}
    \label{fig:mdhp-gds-init-and-tmpt}
\end{figure}

\textit{Standardization} | As illustrated in Figure \ref{fig:mdhp-gds-init-and-tmpt}, a standardization process governed by Equation \ref{eq:mdhp-gds-init-standardize} is applied to every tensor in the input list. This transformation preserves the original distribution shape while scaling the data to a specified range between $Max$ and $Min$, aligns the data across various inputs, ensuring consistency and promoting faster convergence during the training phase.

\begin{equation}\label{eq:mdhp-gds-init-standardize}
    T=\frac{T-min(T)}{max(T)-min(T)}*(Max-Min)+Min
\end{equation}

\textit{Padding} | Although timestamps within each tensor are sorted in ascending order, the tensors often vary in length. This discrepancy in size leads to inefficient data storage and decreased computational performance. To address this problem, we pad each tensor to the length of the longest tensor, \inlinecode{maxTimeLen}, and then stack the tensors into a continuous 2-D array with the shape \inlinecode{(dim, maxTimeLen)}. 

\textit{Decoupling} | A key part of the initialization involves decoupling computations that are unrelated to the gradient descent process. A significant portion of these unrelated calculations is embedded in Part1:
$$
\text{Part1} = \sum_{i=0}^{D-1}\sum_{t=T_i^{(0)}}^{T_i^\text{last}}\text{...}\sum_{j=0}^{D-1}\sum_{k \land T_j^k<t}\text{...}\left(t-T_j^k\right)
$$

This nested calculation poses challenges for two main reasons: (1) ${T_i^\text{last}}$ differs for each dimension, and (2) the constraint $k \land T_j^k < t$ complicates the loop structure. Fortunately, the first issue has been resolved by padding the tensors into $T_\text{padded}$, ensuring uniform tensor lengths. We alleviate the second issue by constructing a 4-D tensor \inlinecode{tMpT} (i.e., $t - T_\text{padded}$) with shape \inlinecode{(dim, maxTimeLen, dim, maxTimeLen)}. When it is necessary, we can easily mask out irrelevant portions from $t - T_j^k$ to streamline the loop and eliminate unwanted calculations. 

\noindent\textbf{+ Stage 2: Part1}

The computational complexity of Part1, as shown in Algorithm \ref{algo:mdhp-gds-persudocode-lnl-part1}, is heightened by its structure, which includes four nested loops and intricate boundary conditions, and this form of loop-heavy computation becomes notably sluggish. However, at the initialization stage, the utilization of $T_\text{padded}$ allows for the pre-computation of the 4-D tensor \inlinecode{tMpT}, which facilitates the computation with underlying vectorized operations and later broadcasting technique.

\textit{Boradcasting} | We employ the vector broadcasting technique to perform vectorized calculations across higher dimensions initially with the help of automatic-broadcasting mechanism in Pytorch \cite{pytorch-braodcasting}, then refining our focus to lower dimensions.

\noindent\textbf{+ Stage 3: Part2}

This is the easiest part to calculate, as shown in Algorithm \ref{algo:mdhp-gds-persudocode-lnl-part2}.

\noindent\textbf{+ Stage 4: Part3}

Similar to Part1, the optimization of Part3 also leverages automatic broadcasting techniques and clamping operations to enhance computational efficiency, as shown in Algorithm \ref{algo:mdhp-gds-persudocode-lnl-part3}. The key distinction for Part3, however, lies in its simpler loop structure. Consequently, the input parameters $\beta$ and $T_\text{padded}$ are extended to a 3-D tensor. This reduction in dimensionality not only decreases the memory usage but also lessens the computational load.

\noindent\textbf{+ Stage 5: JIT Compilation}

Just-In-Time (JIT) compilation is a dynamic method of code compilation that offers a balance between the performance benefits of ahead-of-time (AOT) compilation and the flexibility of interpretation. The principle behind JIT compilation lies in its ability to compile portions of the code during runtime, thereby optimizing the execution speed without requiring a full pre-compilation of the script. 

\subsection{MDHP-NET}
\label{MDHP-NET}

\subsubsection{\textbf{Overall Structure}}

\begin{figure}[!htbp]
    \centering
    \includegraphics[width=0.45\textwidth]{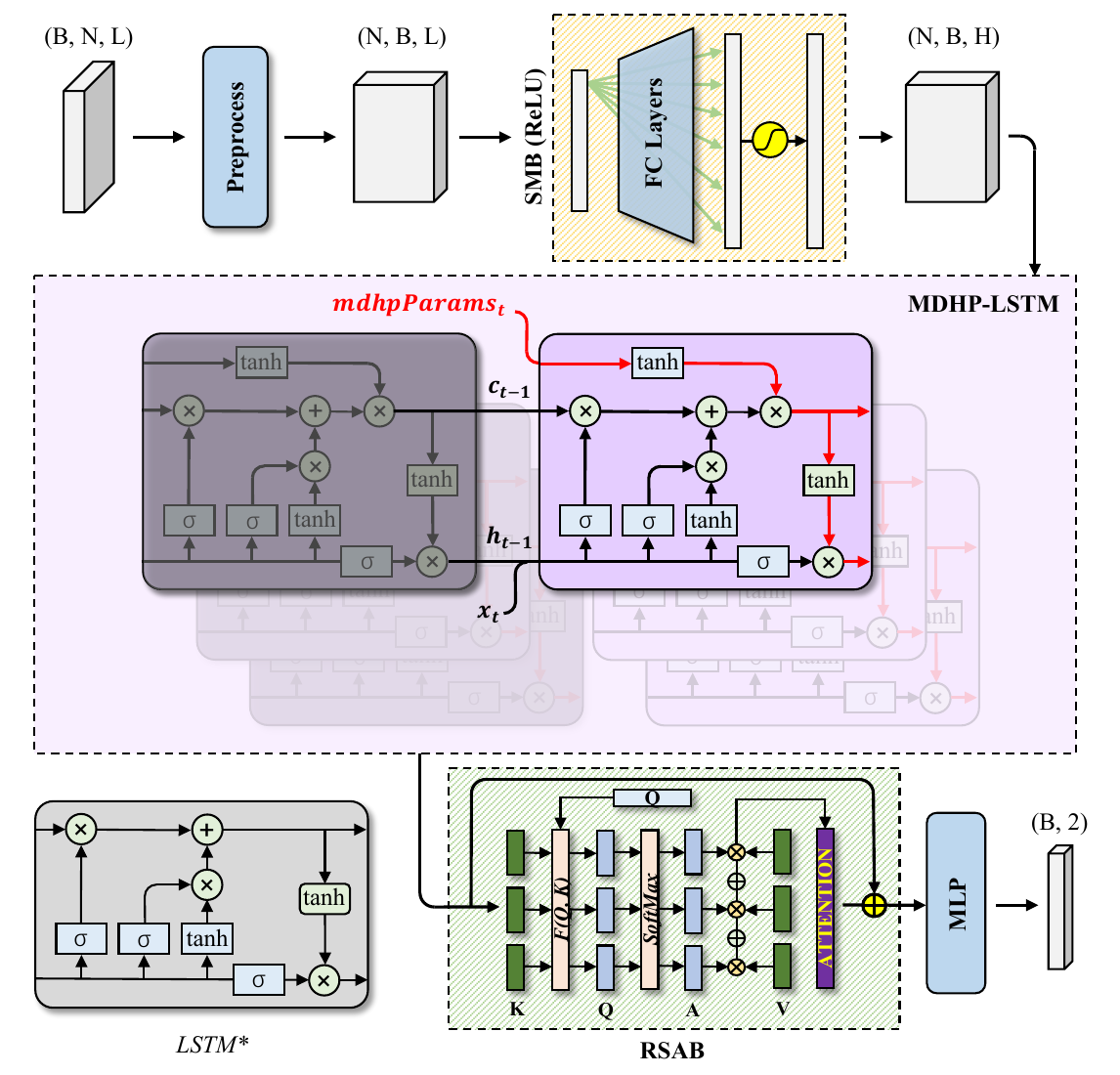}
    \caption{The Overall Structure of MDHP-Net, including \textbf{Preprocessing} block, \textbf{SMB} (Sequence Mapping Block), \textbf{MDHP-LSTM}, \textbf{RSAB} (Residual Attention Block) and a final \textbf{MLP} block. \textbf{B}: Batch Size. \textbf{N}: Number of Sequences per Window. \textbf{L}: Length of Each Sequence. \textbf{H}: Hidden Size. \textbf{*}: LSTM is not used in MDHP-Net, shown here only for comparison with MDHP-LSTM.}
    \label{fig:mdhp-net-structure}
\end{figure}

Figure \ref{fig:mdhp-net-structure} shows the model forwarding process, designed for the analysis of n time-stamped messages within a detection window to perform multiple classification tasks. Within our SOME/IP data described in Section \ref{sec:experiment-dataset}, each message is characterized by 10 specific message fields plus a temporal attribute. 

\subsubsection{\textbf{MDHP-LSTM}}

Traditional LSTM architectures, while effective at capturing temporal dependencies, lack the explicit modeling of event excitation and decay patterns characteristic of cyber attacks. This limitation becomes particularly significant in injection attack detection, where the temporal intensity of events carries crucial discriminative information.

To address this limitation, we propose MDHP-LSTM, which augments the standard LSTM architecture with MDHP parameters as shown in Equation \ref{eq:mdhpnet-mdhp-lstm}:

\begin{equation}
\begin{aligned}
\label{eq:mdhpnet-mdhp-lstm}
\mathbf{i}^{t}&\leftarrow\sigma\left(\mathbf{W}_\mathrm{i}\mathbf{x}^{t-1}+\mathbf{U}_\mathbf{i}\mathbf{h}^{t-1}+\mathbf{b}_\mathbf{i}\right) \\
\mathbf{f}^{t}&\leftarrow\sigma\left(\mathbf{W}_\mathbf{f}\mathbf{x}^{t-1}+\mathbf{U}_\mathbf{f}\mathbf{h}^{t-1}+\mathbf{b}_\mathbf{f}\right) \\
\mathbf{hks}^{t}&\leftarrow\tanh\left(\mathbf{A}\mathbf{\alpha}^\mathrm{x}-\mathbf{B}\left({\beta}^\mathrm{x} \mathbf{T}_\mathrm{span}^\mathrm{x}\right)+\mathbf{C}\mathbf{\theta}^\mathrm{x}\right) \\
\mathbf{\widetilde{c}}^t & \leftarrow \tanh{\left(\mathbf{W}_\mathrm{c}\mathbf{x}^{t-1}+\mathbf{U}_\mathbf{c}\mathbf{h}^{t-1}+\mathbf{b}_\mathbf{c}\right)} \\
\mathbf{c}^{t}&\leftarrow 
\mathbf{hks}^{t} * \left(
    \mathbf{f}^{t} * \mathbf{c}^{t-1} 
    + \mathbf{i}^{t} * \mathbf{\widetilde{c}}
\right) \\
\mathbf{o}^t & \leftarrow \sigma{\left(
    \mathbf{W}_\mathrm{o}\mathbf{x}^{t-1}
    +\mathbf{U}_\mathbf{o}\mathbf{h}^{t-1}
    +\mathbf{b}_\mathbf{o}
\right)} \\
\mathbf{h}^{t}&\leftarrow\mathbf{o}^t * \tanh\left(\mathbf{c}^{t}\right) \\
\mathbf{y}^{t}&\leftarrow\mathbf{W}_\mathrm{y}\mathbf{h}^{t-1}
\end{aligned}
\end{equation}

\setlength{\parindent}{0em}
where $\alpha^x$, $\beta^x$, and $\theta^x$ represent the excitation, decay, and baseline intensity parameters of the Multi-Dimensional Hawkes process, respectively. The matrices $\mathbf{A}$, $\mathbf{B}$, and $\mathbf{C}$ learn to map these parameters to the module's internal state space.
\setlength{\parindent}{1em}

The Hawkes state modulates both the cell state $\mathbf{c}^t$ and hidden state $\mathbf{h}^t$ through multiplicative interactions, as illustrated by the red paths in Figure \ref{fig:mdhp-net-structure}. This integration allows the network to adaptively adjust its memory based on the temporal intensity patterns captured by the Hawkes process parameters, providing a more nuanced representation of attack dynamics.

\section{Experiment}

\subsection{Overview}

\subsubsection{\textbf{Setup}}

The hardware and software specifications of the experiments are detailed in Table \ref{tab:experiment-device-software-specification}.

\begin{table}[!htbp]
	\centering
    \caption{Device and Software Specification}
	\begin{tabular}{cc}
        \hline
		Device / Software & Name / Version                        \\ \hline
		CPU               & Intel Xeon Silver 4210 CPU @ 2.20 GHz \\
		GPU               & NVIDIA GeForce RTX 3090          \\
		System            & Ubuntu 24.04 LTS                      \\
		PyTorch           & 2.6.0+cu126                           \\ \hline
	\end{tabular}
	\label{tab:experiment-device-software-specification}
\end{table}


\subsubsection{\textbf{Evaluation baselines and metrics}}

To rigorously evaluate the proposed MDHP-GDS and MDHP-LSTM, we meticulously crafted a series of experiments tailored to their specific frameworks to confirm their efficacy.

In Section \ref{section:experiment-mdhp-gds-estimation-speed}, we conduct a thorough examination of the estimation speed of MDHP parameters by varying the input data size sent to MDHP-GDS. We executed identical tests on both CPUs and GPUs, demonstrating that our optimizations significantly enhance GPU performance for parameter estimation in large datasets.

In Section \ref{section:experiment-mdhp-gds-estimation-results}, we present evaluations of MDHP-GDS using our simulated SOME/IP communication messages. We employed the Cumulative Distribution Function (CDF) as a visual metric. 






\begin{figure}[htbp]
\centering
\includegraphics[width=0.48\textwidth]{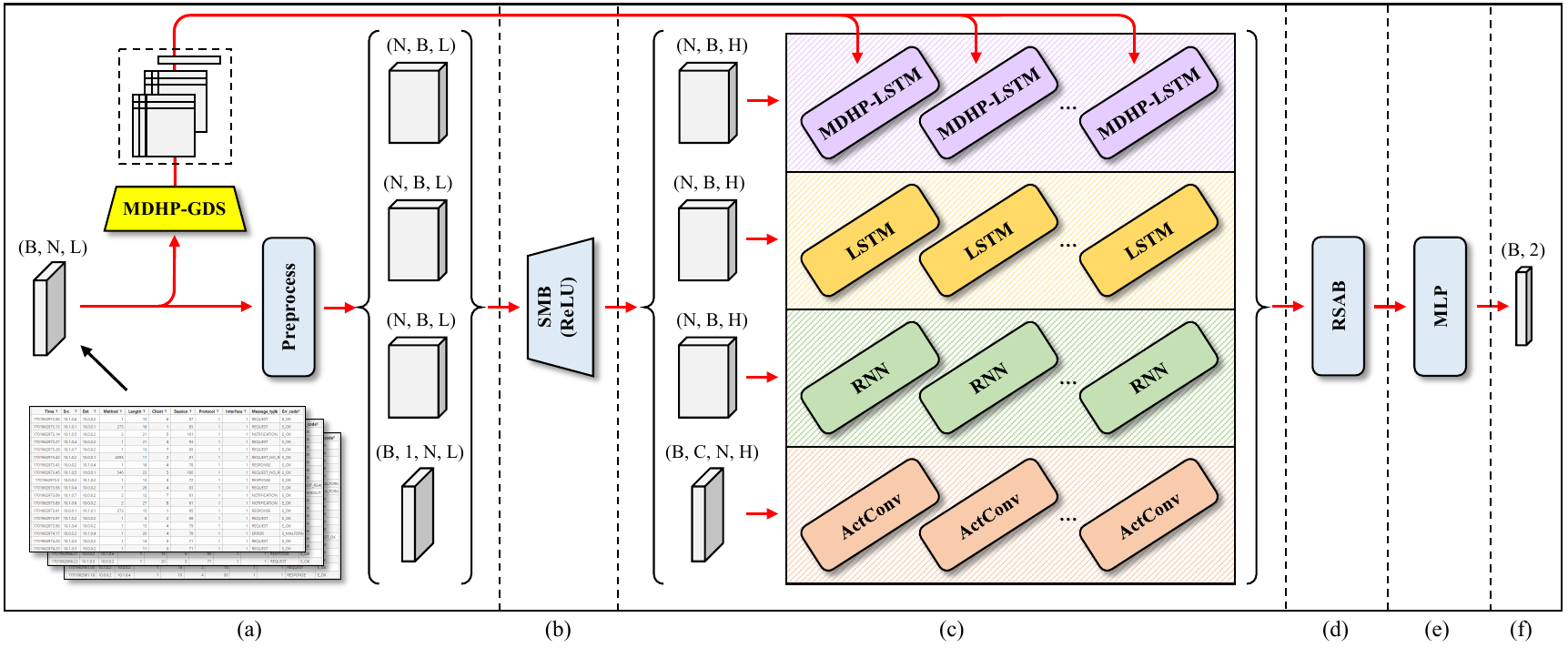}
\caption{Contrast and Ablation Experiment of MDHP-Net. We compared four different models, MDHP-Net, SISSA-LSTM, SISSA-RNN and SISSA-CNN, by keeping (a), (b), (d), (e) and (f) the same, while changing the backbone blocks in (c). Moreover, MDHP-Net (with MDHP-LSTM blocks) and SISSA-LSTM (with LSTM blocks) comprise the ablation experiment on MDHP parameters.}
\label{fig:experiment-mdhp-net-contrast-and-ablation}
\end{figure}

In Section \ref{section:experiment-mdhp-net-contrast-and-ablation-experiment}, we designed extensive comparative and ablation experiments to demonstrate the significant advantages of MDHP-Net in detecting injection attacks on in-vehicle SOME/IP communication messages. As illustrated in Figure \ref{fig:experiment-mdhp-net-contrast-and-ablation}, we compared MDHP-Net with three models with different architectures proposed in SISSA\cite{liu2024sissa}. To ensure the fairness of the experiments, we maintained the overall structure of the models while replacing the key module in Figure \ref{fig:experiment-mdhp-net-contrast-and-ablation}-(c) with multiple MDHP-LSTM blocks to complete the construction of MDHP-Net. In general, we aimed to compare MDHP-Net with SISSA-LSTM, SISSA-RNN, and SISSA-CNN under conditions that ensure as much fairness as possible.

\subsection{Dataset}\label{sec:experiment-dataset}

\subsubsection{\textbf{STEIA9}}

We proposed a new dataset, \textbf{STEIA9} (SOME/IP Time-Exciting Injection Attack under 9 Patterns), to evaluate the performance of MDHP-GDS and MDHP-Net in the experiment. The dataset is carefully designed for simulating the time-exciting attack patterns which are difficult for normal models to detect, and has been opened to public\footnote{\url{https://www.dropbox.com/scl/fi/wjl7z09dx8z7md5p24hk4/STEIA9.7z?rlkey=6tb4zoxll4dhtt1x7gkddo71m&st=rpvvo1mi&dl=0}}.


The generation of STEIA9 utilizes a suite of open-source tools from Github, including the SOME/IP-Generator\footnote{\url{https://github.com/Egomania/SOME-IP_Generator}}, SISSA\footnote{\url{https://github.com/jamesnulliu/SISSA}}, and MDHP-NET\footnote{\url{https://github.com/Tiara8735/MDHP-Net-Anonymous/tree/main}}. Additionally, we modified the source code of SISSA and applied our time-exciting threat model in Section \ref{sec:threat-model} to the simulation process. The attack implementations are detailed in section \ref{sec:attack-rate-function}.

As shown in Table \ref{tab:experiment-dataset-train-and-val}, STEIA9 was randomly partitioned into training and validation sets with an 8:2 split, maintaining equal proportions of normal and attack windows to prevent class imbalance. The model weights were frozen after training, and no validation set data was used for model fine-tuning, ensuring unbiased evaluation of the model's performance.

\begin{table}[!htbp]
	\centering
    \caption{Breif Information of STEIA9. Each data is a message window including 128 consecutive SOME/IP packets. Each window is labeled as Normal or Attack. The explanation of Attk Rate, IP Ctrl and Sample is in the following 2 sections.}
	\begin{tabular}{cccccc}
		\hline
		ID & Train (x128) & Val (x128) & Attk Rate & IP Ctrl & Sample \\ \hline
		00 & 8282             & 2072           & PLA         & PLA        & NPP    \\
		01 & 9444             & 2362           & DEA         & DEA        & NPP    \\
		02 & 10960            & 2740           & ASA         & ASA        & NPP    \\
		03 & 10636            & 2660           & DAM         & DAM        & NPP    \\
		04 & 10716            & 2680           & /           & DAM        & DRP    \\
		05 & 8160             & 2040           & PLA         & PLA        & ND     \\
		06 & 9416             & 2356           & DEA         & DEA        & ND     \\
		07 & 10976            & 2746           & ASA         & ASA        & ND     \\
		08 & 10420            & 2606           & DAM         & DAM        & ND     \\ \hline
	\end{tabular}
	\label{tab:experiment-dataset-train-and-val}
\end{table}

\subsubsection{\textbf{Simulating Attack Execution}}\label{sec:attack-rate-function}

To simulate the process of time-exciting injection attacks in an in-vehicle Ethernet environment based on the SOME/IP protocol, we conduct a series of simulation experiments. In contrast to real-world CAN bus attacks in section \ref{sec:threat-model}, we assume that a single valid message is injected at each injection time point. The specific method is as follows:

Firstly, the attack behavior is governed by two key parameters: the number of controlled source IPs \( n_{\text{victims}} \), and the probability of injecting a valid message at a given time, denoted by \( \text{inject\_prob} \). Based on four realistic and feasible attack strategies, we construct a corresponding injection rate function \( \text{inject\_rate} = g(t) \) and a controlled IP function \( n_{\text{victims}} = f(t) \), which describe the temporal evolution of the attack intensity and the number of IPs under the attacker's control, respectively. Secondly, after determining the forms of the rate function \( g(t) \) and the IP control function \( f(t) \), we normalize the timestamps of the window (containing 128 × 3 messages) to map them into the interval \([0, 1]\). Within this normalized time domain, and to control the temporal distribution of attack events, we introduce a thinning process. Specifically, we construct three types of sampling functions to generate candidate attack time points. These candidates are then filtered using an acceptance-rejection sampling method, where each candidate is accepted with probability \( g(t)/g_{\text{max}} \), resulting in the final set of injection event times. Finally, for each injection event time, the current value of \( f(t) \) determines the number of source IPs the attacker can control at that moment. One IP is randomly selected from this pool as the source address, while the destination IP is randomly chosen from all available valid IPs to inject a single valid message.

In our simulation, probability is used in place of intensity (determined by $g(t))$ to reflect the attacker’s injection strategy. The injection probability \( \text{inject\_prob}\) is proportional to the attack intensity. This design still effectively simulates the original strategy, as higher intensity corresponds to a higher probability of injection. The four attack rate functions used in our simulation correspond to the strategies listed in Table \ref{tab:Attack strategy}, and their associated IP control functions are detailed in Appendices \ref{appendices_IP_control}. Meanwhile, sampling functions are detailed in Appendices \ref{algo:appendices_sampling_algorithm}.


\subsection{MDHP-GDS: Estimation Speed}
\label{section:experiment-mdhp-gds-estimation-speed}

We evaluated the estimation speed of  MDHP-GDS using two quantitative metrics.

\begin{itemize}[topsep=0pt, left=0pt]
    \item \textbf{Window Time}: Seconds spent for estimating the MDHP parameters for one window in average.
    \item \textbf{Throughput}: Number of messages MDHP-GDS can handle per second. In our experiment, number of messages is randomly generated according to the test configurations but recorded precisely after generation.
\end{itemize}

For experimental reproducibility, we set the \inlinecode{Random Seed} to 2024 and enable both \inlinecode{torch.compile} and optimized log-likelihood computation. 

As mentioned in previous sections, \inlinecode{torch.compile} implements Just-In-Time (JIT) compilation and kernel-fusion instead of interpreted execution, which is consistently enabled being fair across all experiments. The optimized log-likelihood computation is crucial, as the unoptimized version exhibits prohibitive computational overhead even for small-scale datasets. 


\begin{table}[!htbp]
	\centering \caption{Estimation Speed of MDHP-GDS with CPU.}
	\begin{tabular}{ccccc}
		\hline
		Dim & Max-T-Len & Min-T-Len & Window-Cost & Throughput \\ \hline
		5        & 10        & 5         & 0.3144      & 4.3255     \\
		10       & 100       & 50        & 0.4988      & 61.7397    \\
		15       & 100       & 50        & 0.5982      & 74.6160    \\
		20       & 100       & 50        & 0.6896      & 88.7971    \\
		25       & 100       & 50        & 0.9819      & 72.8373    \\
		30       & 100       & 50        & 1.3888      & 61.5767    \\
		30       & 150       & 50        & 2.0903      & 61.3496    \\
		30       & 200       & 50        & 3.4459      & 42.4848    \\ \hline
	\end{tabular}
	\label{tab:experiment-mdhp-gds-estimation-speed-cpu}
\end{table}

Table \ref{tab:experiment-mdhp-gds-estimation-speed-cpu} presents the parameter estimation performance of MDHP-GDS in CPU mode. The results demonstrate that as the input scale increases, the Window-Cost exhibits a clear upward trend, while throughput peaks at 88.7971 packets per second before declining. This performance characteristic can be attributed to our optimization algorithm's efficient utilization of CPU vector operations and parallel processing capabilities.

\begin{table}[!htbp]
	\centering \caption{Estimation Speed of MDHP-GDS with GPU.}
	\begin{tabular}{ccccc}
		\hline
		Dim & Max-T-Len & Min-T-Len & Window-Cost & Throughput \\ \hline
		5        & 10        & 5         & 0.4079      & 3.3335     \\
		10       & 100       & 50        & 0.5710      & 53.9324    \\
		15       & 100       & 50        & 0.4807      & 92.8501    \\
		20       & 100       & 50        & 0.3383      & 181.0082   \\
		25       & 100       & 50        & 0.3902      & 183.2755   \\
		30       & 100       & 50        & 0.3471      & 246.3298   \\
		30       & 150       & 50        & 0.3173      & 404.0592   \\
		30       & 200       & 50        & 0.3469      & 421.9364   \\ \hline
	\end{tabular}
	\label{tab:experiment-mdhp-gds-estimation-speed-gpu}
\end{table}

Table \ref{tab:experiment-mdhp-gds-estimation-speed-gpu} illustrates the parameter estimation performance of MDHP-GDS in GPU mode. While GPU mode shows marginally lower performance than CPU mode for small-scale datasets, it demonstrates superior scalability as data dimensions and timestamp volumes increase. This superior scaling behavior can be attributed to the synergy between our vectorized computation optimizations and the GPU's massive parallel threading architecture. For high-dimensional Hawkes process parameter estimation tasks, we recommend the GPU implementation of MDHP-GDS for optimal estimation performance.

\subsection{MDHP-GDS: Estimation Results}
\label{section:experiment-mdhp-gds-estimation-results}

\begin{figure}[!htbp]
    \centering
    \includegraphics[width=0.5\textwidth]{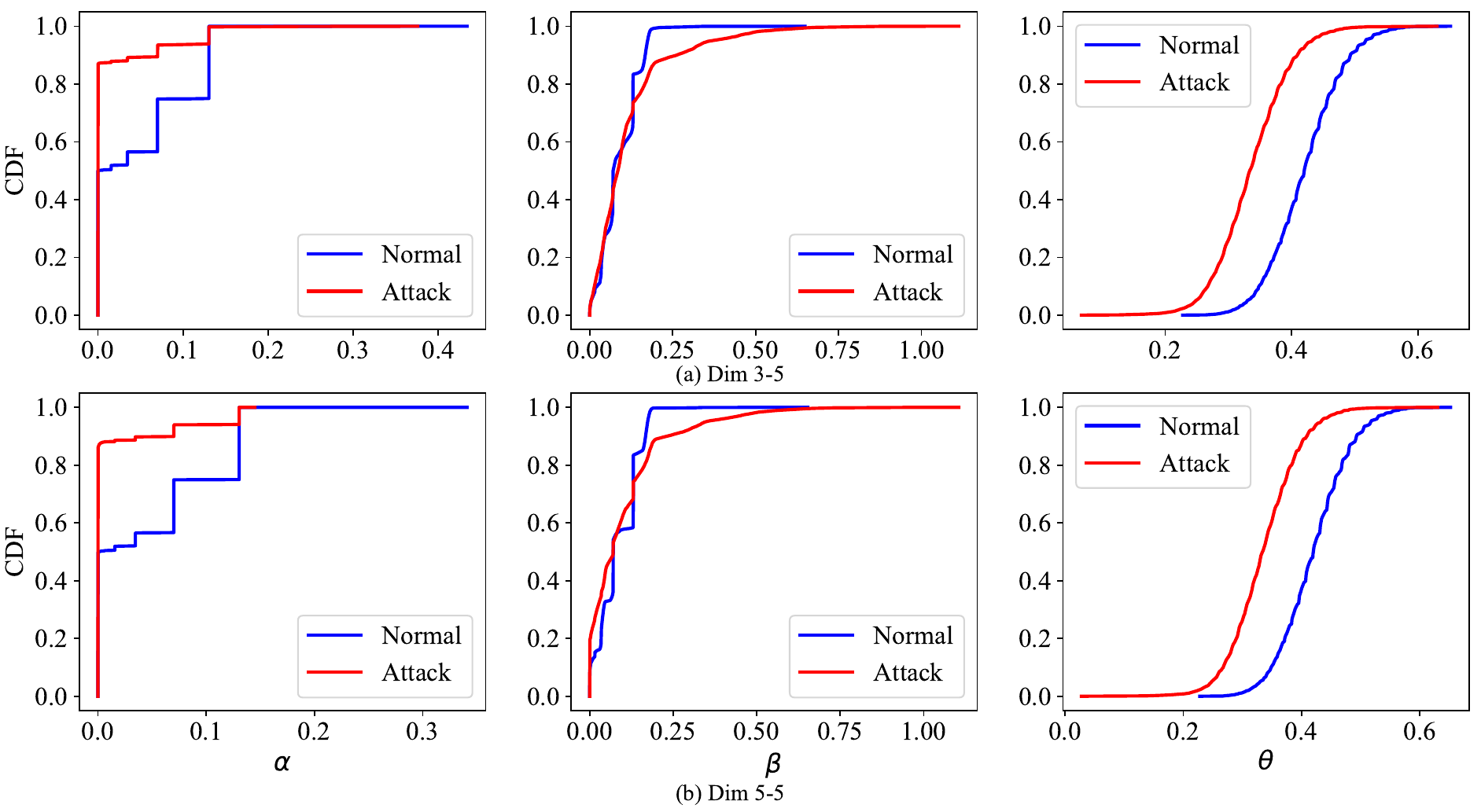}
    \caption{CDF of (a) Dim 3-5 and (b) Dim 5-5.}
    \label{fig:experiment-cdf-3-5-and-5-5}
\end{figure}
We evaluated the MDHP parameter estimation performance of our MDHP-GDS framework using simulated SOME/IP communication data. For each temporal window, the MDHP parameters consist of three components: the excitation matrix $\alpha \in \mathbf{R}^{Dim \times Dim}$, the decay matrix $\beta \in \mathbf{R}^{Dim \times Dim}$, and the baseline intensity vector $\theta \in \mathbf{R}^{Dim}$, where $Dim$ denotes the number of ECUs in the system.

Figure \ref{fig:experiment-cdf-3-5-and-5-5} illustrate the CDF plots for two representative inter-ECU communications: Dim 3-5 (communication from ECU 3 to ECU 5) and Dim 5-5 (self-exciting process of ECU 5). The analysis reveals distinct parameter distributions between normal and attack scenarios, providing discriminative features for the subsequent deep learning process. Additional CDF plots for other dimensions can be found in Appendix \ref{Appendices_CDF}.



These distinctive parameter distributions quantify the temporal perturbations in SOME/IP communication patterns during injection attacks. The clear separability between normal and attack distributions validates our approach and motivates our proposed MDHP-Net architecture, which leverages these temporal dynamics for attack detection. The ability to capture fine-grained differences in inter-ECU communication patterns through MDHP parameter estimation provides a robust foundation for automotive network intrusion detection.

\subsection{MDHP-Net: Constrast and Ablation Experiment}
\label{section:experiment-mdhp-net-contrast-and-ablation-experiment}

\subsubsection{\textbf{Performance in Training}}

\begin{figure*}[!htbp]
\centering
\includegraphics[width=0.95\linewidth]{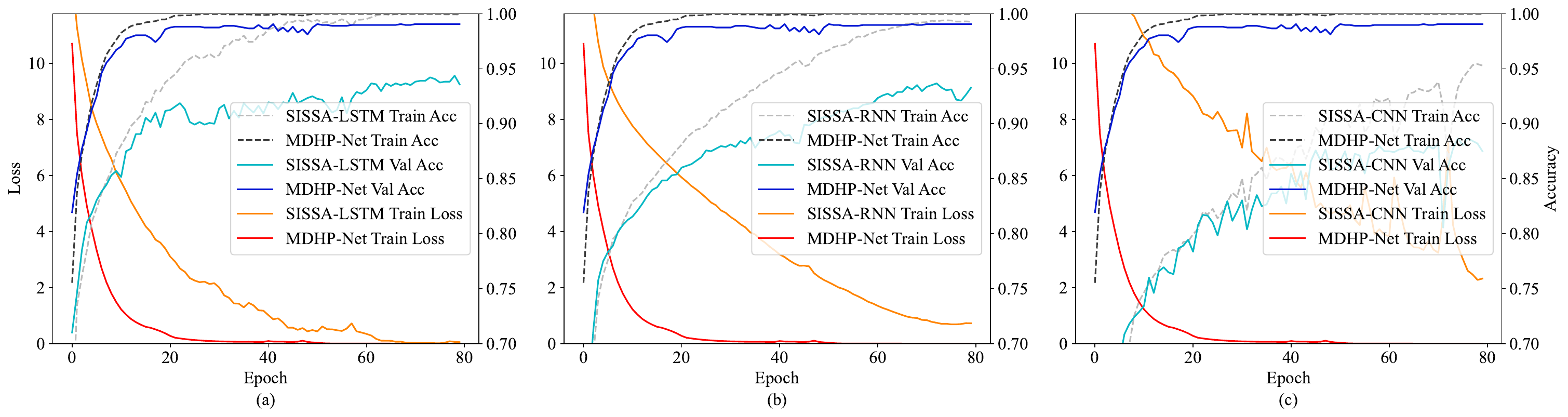}
\caption{Train, Val Accuracy and Train Loss of MDHP-Net Compared with (a) SISSA-LSTM, (b) SISSA-RNN, (c) SISSA-CNN on STEIA9-03.}
\label{fig:experiment-acc-loss}
\end{figure*}

Figure \ref{fig:experiment-acc-loss} illustrates the comparative training dynamics on STEIA9-03 of MDHP-Net compared to the baseline models. In subfigures (a) and (b), MDHP-Net, SISSA-RNN, and SISSA-LSTM demonstrate rapid initial loss reduction in the training set, indicating effective feature capture from the SOME/IP message windows. Among all models, MDHP-Net exhibits notably faster convergence. During epochs 60-80, while SISSA-RNN and SISSA-LSTM almost achieve 0.99 accuracy in the training set, their validation performance remains substantially lower, indicating non-negligible overfitting. In contrast, MDHP-Net demonstrates better generalization, with closer alignment between training and validation accuracies, suggesting enhanced model robustness. 

We attribute this improved performance to the MDHP parameters providing meaningful temporal feature information during inference, enabling better discrimination between normal and anomalous patterns, which aligns with our initial hypothesis.

In general, the experimental results demonstrate MDHP-Net's superior performance across several metrics:

\begin{itemize}[topsep=0pt, left=0pt]
\item Faster convergence compared to SISSA-LSTM and SISSA-RNN.
\item Enhanced training stability.
\item Superior validation set performance - Highest validation accuracy of MDHP-Net: 0.989; of SISSA-LSTM: 0.938; of SISSA-RNN: 0.936; of SISSA-CNN: 0.882.
\item While SISSA-CNN achieves high training accuracy, its significant underfitting in the validation set indicates inadequate temporal feature capture, confirming the superiority of temporal neural architectures for this specific intrusion detection scenario.
\end{itemize}

Table \ref{tab:experiment-hyperperameters} presents the training hyperparameters. To ensure reproducibility in our distributed training setup, we initialized the random seed as 1024 + RANK, where RANK represents the process identifier. For MDHP-Net and sequence-based models (SISSA-LSTM, SISSA-RNN), we used lower learning rate and weight decay values ($5e^{-5}$), while SISSA-CNN required higher values ($3e^{-4}$) to achieve effective feature learning.

\begin{table}[htbp]
	\centering
 	\caption{Hyperparameters of Traning.}
	\begin{tabular}{cc}
		\hline
		Hyperparamerter & Value       \\ \hline
		Random Seed     & 1024 + RANK \\
		Max Epoch       & 50          \\
		Optimizer       & AdamW       \\
		Learning Rate   & $5e^{-5}$ / $3e^{-4}$ \\
		Weight Decay    & $5e^{-5}$ / $3e^{-4}$ \\ \hline
	\end{tabular}
	\label{tab:experiment-hyperperameters}
\end{table}

\subsubsection{\textbf{Metrics}}

\begin{figure}[H]
    \centering
    \includegraphics[width=0.45\textwidth]{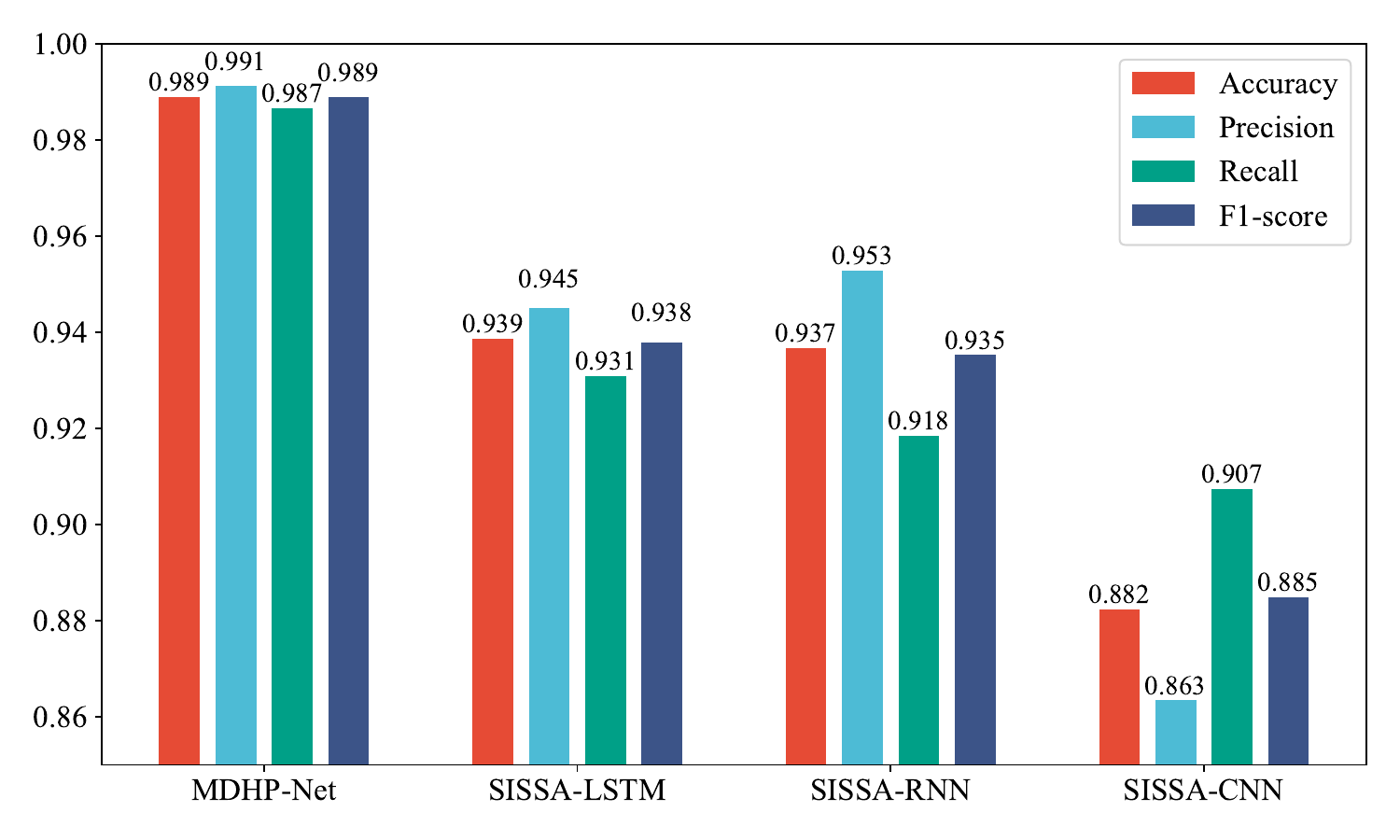}
    \caption{Comparison of Metrics on STEIA9-03.}
    \label{fig:experiment-metrics}
\end{figure}

Figure \ref{fig:experiment-metrics} presents a comparative performance evaluation of the proposed MDHP-Net against three baseline models (SISSA-CNN, SISSA-RNN, and SISSA-LSTM) on STEIA9-03 using standard metrics: accuracy, precision, recall, and F1-score. MDHP-Net demonstrates superior performance across all metrics.

\begin{figure*}[!htbp]
    \centering
    \includegraphics[width=0.83\textwidth]{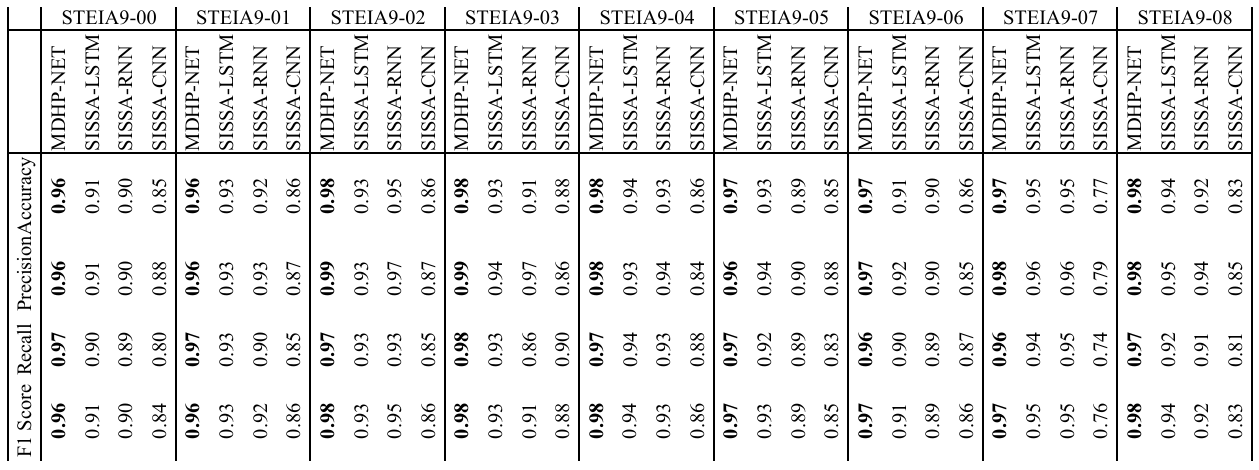}
    \caption{Full Test Results of 4 Models on Complete STEIA9.}
    \label{fig:full-test-results}
\end{figure*}

Figure \ref{fig:full-test-results} shows the full test results of the four models in complete STEIA9. Our proposed model, MDHP-Net, achieved the highest accuracy, precision, recall, and F1 score of identifying the attack windows under all different attack patterns. 



\begin{figure}[H]
    \centering
    \includegraphics[width=0.47\textwidth]{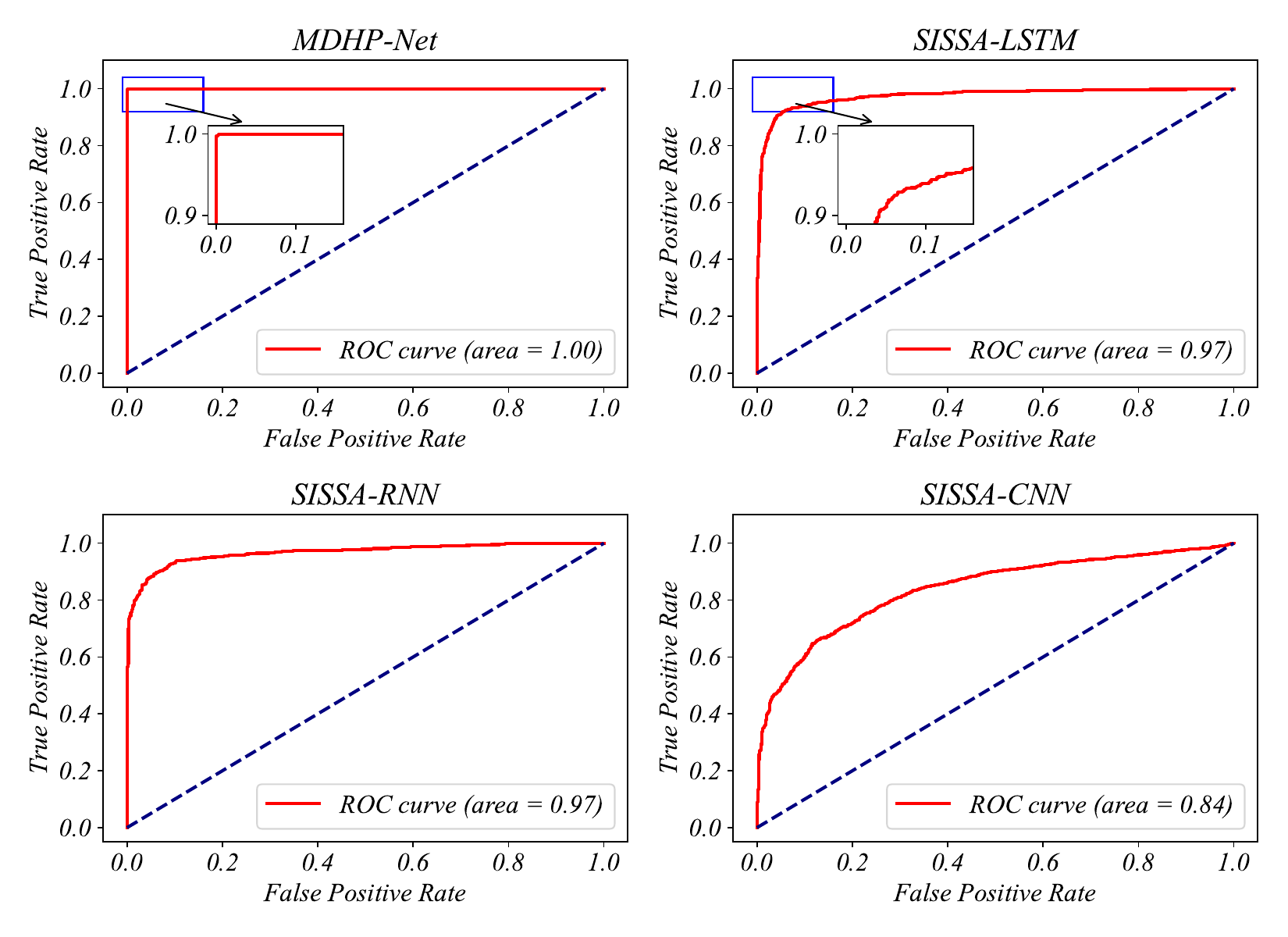}
    \caption{ROC and AUC of Different Models.}
    \label{fig:experiment-roc-cur}
\end{figure}

Figure \ref{fig:experiment-roc-cur} presents the ROC curves and corresponding AUC metrics for all models. MDHP-Net and SISSA-LSTM achieve optimal AUC scores of 1.00, demonstrating perfect discriminative capability across all classification thresholds. The magnified inset of their ROC curves in the low false positive rate region (0-0.1) reveals consistently high true positive rates, crucial for practical deployment where false alarms must be minimized. 


All these enhancements can be attributed to the MDHP-LSTM module, which incorporates MDHP parameters to model both excitation and decay patterns in the temporal domain. In contrast, all baseline models exhibit limitations in capturing the temporal dynamics of injection attacks.





\subsection{MDHP-Net: Computational Cost Analysis}

Table \ref{tab:experiment-comp-cost} presents a comparative analysis of model complexity and computational requirements. MDHP-Net has approximately $5.5\%$ more parameters (2.55M) than SISSA-LSTM (2.41M) due to the additional Hawkes process components. Despite this increased parametric complexity, MDHP-Net maintains comparable memory efficiency, requiring only $18.6\%$ more total memory (113.03MB vs. 95.32MB) than SISSA-LSTM. SISSA-CNN, while having the smallest parameter count (303K), demonstrates significantly higher computational overhead with a forward/backward pass size of 2839.68MB - approximately 28 times larger than MDHP-Net (100.80MB). In contrast, the sequence-based architectures (MDHP-Net, SISSA-RNN, and SISSA-LSTM) exhibit more efficient memory utilization due to their sequential processing nature.
\begin{table}[htbp]
	\centering
    \caption{Computational Cost of Different Models}
	\begin{tabular}{ccccc}
		\hline
 \parbox[c][2em][c]{2cm}{\centering Model Name}
& \parbox[c][2em][c]{1.5cm}{\centering MDHP-Net} & \parbox[c][2em][c]{1cm}{\centering SISSA-LSTM} & \parbox[c][2em][c]{1cm}{\centering SISSA-RNN} & \parbox[c][2em][c]{1cm}{\centering SISSA-CNN} \\ \hline
		Total Params             & 2548738  & 2414594    & 2216450   & 303459    \\
		Input Size            & 2.04     & 1.64       & 1.64      & 1.64      \\
		F/B Pass Size   & 100.80   & 84.02      & 84.02     & 2839.68   \\
		Params Size           & 10.19    & 9.66       & 8.87      & 1.21      \\
		Total Size  & 113.03   & 95.32      & 94.52     & 2842.53   \\ \hline
	\end{tabular}
\caption*{\footnotesize All sizes are specified in MB (except for Total Params)}
\label{tab:experiment-comp-cost}
\end{table}

\section{Discussion and Limitations}

\textbf{Towards Secure Automotive Ethernet.} SOME/IP is an automotive protocol addressing bandwidth limitations of traditional vehicle networks like CAN bus, MOST, and FlexRay. However, vulnerabilities in commercial vehicle network stacks, especially in Bluetooth, Wi-Fi, and 4G modules, increase the risk of unauthorized remote access to IVN\cite{miller2015remote}\cite{keenlab2018}. To support research on emerging threats, we introduce the first open-source automotive Ethernet dataset with diverse time-exciting attack patterns, establishing a benchmark for IVN security research.\\
\textbf{Further Optimization of MDHP-GDS.} Although the current implementation benefits from vectorization and GPU acceleration, further improvements are possible. Processing observation windows in batches using tensor padding and alignment can better leverage GPU parallelism and memory bandwidth. Additionally, reimplementing the core solver in C++ with custom CUDA kernels can reduce computational latency and optimize memory access, especially for gradient computations.\\
\textbf{Architectural Design Considerations.} Transformer models excel in NLP and computer vision but face challenges in intrusion detection due to quadratic complexity and high memory demands. MDHP-Net prioritizes efficiency by using lightweight MDHP-LSTM blocks and a single RSAB module, balancing accuracy and resource use for real-time vehicular network security.

\section{Related work}
\textbf{Injection attack detection for CAVs}. The current focus is on detecting and mitigating various data injection attacks targeting connected vehicles, particularly emphasizing the security of the CAN and in-vehicle communication systems. Ben et.al \cite{ben2020performance} evaluates multiple CAN bus attack detection methods, highlighting the effectiveness of Pearson correlation and noting the high false positive rates associated with unsupervised learning techniques, thereby underscoring the importance of incorporating knowledge about ECU collaboration. Building upon this foundation, Ji et.al \cite{ji2024vehicle}introduce an enhanced CNN algorithm designed to detect abnormal data in in-vehicle networks, effectively addressing high false negative rates through the utilization of data field characteristics and Hamming distance. Jedh et.al \cite{jedh2021detection}propose a method that utilizes message sequence graphs and cosine similarity to enhance real-time cybersecurity against message injection attacks on the CAN bus. Biroon et.al \cite{biroon2021false} develop a monitoring system that employs a partial differential equation model to detect and isolate false data injection attacks within Cooperative Adaptive Cruise Control platoons, thereby enhancing safety and reliability. In contrast, Siddiqi et.al \cite{siddiqi2024multichain} introduce a multichain framework employing decentralized blockchain technology to enhance security against false data injection attacks in connected autonomous vehicles, thereby reducing reliance on cloud services and improving transaction throughput. Zhao et.al \cite{zhao2021detection} present a cloud-based sandboxing framework designed to detect false data injection attacks in connected and automated vehicles, employing isolation techniques to enhance the safety of V2X communication systems. Nonetheless, these works elevated false positive and negative rates, complexity in real-time applications, and reliance on computational resources. They cannot identify the injected abnormal messages with normal traffic patterns, especially for time-excitation injection attacks.\\
\textbf{Parameter estimation based on the Hawkes process}. 
Parameter estimation for the intensity function is a crucial aspect of modeling with the Hawkes process. Pan et al.\cite{pan2022quantifying} propose a multistage Hawkes process and used the Expectation-Maximum (EM) algorithm to solve for the Hawkes parameters. Hridoy Sankar Dutta et al.\cite{dutta2020hawkeseye} use the Hawkes process to address the problem of identifying fake retweeters, applying methods such as gradient descent to maximize the log-likelihood function for parameter estimation. For faster convergence, Yao et al.\cite{yao2021stimuli} employ an accelerated gradient method framework to perform parameter inference in their proposed stimuli-sensitive Hawkes process model. In practice, traditional methods such as gradient descent and the EM algorithm become challenging to compute for multidimensional Hawkes processes due to their parameter complexity, leading researchers to explore more efficient algorithms. R´emi Lemonnier\cite{lemonnier2017multivariate} address this issue by developing an iterative algorithm based on adaptive matrix projection and self-concordant convex optimization, tackling the difficulties of parameter estimation in multivariate Hawkes processes. Additionally, Mei\cite{mei2017neural} apply Monte Carlo sampling to estimate gradients and optimize the event sampling process, enabling parameter estimation for multivariate Hawkes processes.

\section{Conclusion}
In this paper, we present a novel threat model targeting IVNs: time-exciting message injection attacks, which gradually manipulate network traffic to disrupt vehicle operations and compromise safety-critical functions. We first conduct a security analysis of this threat model, characterizing its key properties—dynamic behavior, time-exciting escalation, and limited prior knowledge. To validate the attack’s feasibility, we replicate it on a real ADAS system via the CAN bus, exploiting UDS service vulnerabilities and proposing four attack strategies. We further investigate its applicability in Ethernet-based SOME/IP networks. To counter such threats, we propose a novel detection method for time-exciting injection attacks. Additionally, we release the first open-source dataset for time-exciting attacks, covering nine Ethernet-based attack scenarios. Experimental results demonstrate that our MDHP-GDS achieves high-speed estimation and effectively discriminates between different attack scenarios. Moreover, MDHP-Net exhibits strong detection performance, outperforming baseline methods in identifying injection attacks on IVNs.









\printbibliography

\section{Appendices}
\subsection{Likelihood function}\label{Appdenix A}
In Section \ref{mdhp-gds-likelihood}, we present the structure of the likelihood function for the Multi-Dimensional Hawkes process. We will now provide the full derivation of that equation, repeated here:
\begin{equation}\label{eq:L}
 L = \prod_{i=0}^{D-1} \prod_{t=T_i^{(0)}}^{T_i^{\text{last}}} {\lambda}^i(t) \cdot e^{-\int_0^{T_{\text{span}}} {\lambda}^i(v) dv} \tag{3}
\end{equation}

Firstly, building on the foundation laid in \cite{ozaki1979maximum}, we provide a more complete derivation of the univariate Hawkes process. Equation \ref{eq:intensity function-Hawkes process} provides the intensity function for a univariate Hawkes process. Since the Hawkes process is a type of regular point process, we redefine the intensity function based on the definition of point processes \cite{daley2003introduction}:
\begin{equation}\label{eq:intensity function-3a}
\lambda(t)=\frac{f(t|\mathcal{F}_{t_{n}})}{1-\int_{t_{n}}^t{f(t|\mathcal{F}_{t_{n}})}} \tag{3a}
\end{equation}
where \( f(t| \mathcal{F}_{t_{n}}) \) is the conditional probability density function. 

To calculate the likelihood function, we need to determine \( f(t| \mathcal{F}_{t_{n}}) \). For Equation \ref{eq:intensity function-3a}, we can further transform it as follows \cite{rasmussen2018lecture}:
$$\begin{aligned}\lambda(t)&=\frac{\frac{d}{dt}\mathrm{F}\left(t|\mathcal{F}_{{t}_{n}}\right)}{1-\mathrm{F}\left(t|\mathcal{F}_{t_{n}}\right)}\\&=-\frac{d}{dt}\ln(1-\mathrm{F}\left(t|\mathcal{F}_{t_{n}}\right))\end{aligned}$$

Integrate both sides with respect to \( t \):

\begin{equation}\label{eq:intensity function-3b}
    \int_{t_{n}}^t\lambda(v)dv=-\ln(1-\mathrm{F}\left(t|H_{t_{n}}\right)) \tag{3b}
\end{equation}

So that,
\begin{equation}\label{eq:intensity function-3c}
    1-\mathrm{F}\left(t|\mathcal{F}_{t_{n}}\right)=e^{-\int_{t_{n}}^t\lambda(v)dv} \tag{3c}
\end{equation}

We get the \( f(t|\mathcal{F}_{t_{n}}) \) by substituting Equation \ref{eq:intensity function-3c} into Equation \ref{eq:intensity function-3a}:

\begin{equation}{f(t|\mathcal{F}_{t_{n}})=\lambda(t)\cdot}e^{-\int_{t_{n}}^t\lambda(v)dv} \tag{3d}
\end{equation}

The likelihood function expression is given by the following equation:
$$ L=f\left(t_1\mid\mathcal{F}_0\right)f\left(t_2\mid\mathcal{F}_{t_1}\right)\cdots f\left(t_n\mid\mathcal{F}_{t_{n-1}}\right)\left(1-\mathrm{F}\left(T\mid\mathcal{F}_{t_n}\right)\right)$$
where \(1-\mathrm{F}\left(T\mid\mathcal{F}_{t_n}\right)\) is used to denote the time of the next event \( t_{n+1} \) in the Hawkes process, which is greater than \( T \).

So, we can get the likelihood of univariate Hawkes process:
\begin{equation}
\prod_{i=1}^n\lambda\left(t_i\right)\cdot e^{-\int_0^T\lambda(v)dv} \tag{3e}
\end{equation}

After getting the likelihood function for the one-dimensional Hawkes process, we extend our analysis to the Multi-Dimensional Hawkes process. Here, we introduce the concept of conditional independence:\\
\textbf{\textit{Definition 6:}} Conditional independence implies that, given the entire history, event sequences across different dimensions are mutually independent.

For random variables that satisfy conditional independence, there is an important property: their joint probability can be represented as the product of their local conditional probabilities. Clearly, the Multi-Dimensional Hawkes process satisfies conditional independence. The joint likelihood function for a Multi-Dimensional Hawkes process can be expressed as the product of the likelihoods for each dimension. Therefore, we can derive the likelihood function for the Multi-Dimensional Hawkes process as follows:

\begin{equation}
 L = \prod_{i=0}^{D-1} \prod_{t=T_i^{(0)}}^{T_i^{\text{last}}} {\lambda}^i(t) \cdot e^{-\int_0^{T_{\text{span}}} {\lambda}^i(v) dv} \tag{3f}
\end{equation}

\subsection{Simplifications for log-likelihood function}\label{Appdenix B}

For the proposed log-likelihood in Equation \ref{eq:log-likelihood}, we define the integral part as \(\Gamma\) :

\begin{equation}
    \Gamma=\sum_{i=0}^{D-1} \int_{0}^{T_{\text{span}}} \left( \theta_i+\sum_{j=0}^{D-1} \sum_{k : T_j^k < v} \alpha^{(i,j)} e^{-\beta^{(i,j)} (v-T_j^k ) } \right) dv
\end{equation}

For \(\Gamma\) , we can further solve the integral to obtain a more simplified form. First, for the baseline intensity \(\theta\), based on the definition of the integral, we can isolate it:
\begin{equation}
    \sum_{i=0}^{D-1} \int_0^{T_{\text{span}}}  \theta_i \, dv =T_{\text{span}}\sum_{i=0}^{D-1}\theta_i
\end{equation}

Thus, we define the remaining part as \(\Gamma'\), and further derive the following for \(\Gamma'\):

\begin{align}\label{eq:log-likelihood-derivation process}
    \Gamma' &= \sum_{i=0}^{D-1} \sum_{j=0}^{D-1} \int_0^{T_{\text{span}}}  \left( \sum_{k \text{ such that } T_j^k < t} \alpha^{(i,j)} e^{-\beta^{(i,j)} (v-T_j^k ) } \right) dv \notag \\
    &= \sum_{i=0}^{D-1} \sum_{j=0}^{D-1} \alpha^{(i,j)} \int_{T_j^{\text{last}}}^{T_{\text{span}}}  \left( \sum_{k=T_j^1}^{T_j^{\text{last}}} e^{-\beta^{(i,j)} (v-T_j^k ) } \right) dv + \notag \\
    &\quad \sum_{i=0}^{D-1} \sum_{j=0}^{D-1} \alpha^{(i,j)}  \sum_{z=1}^{\text{last}_j-1} \int_{T_j^z}^{T_j^{z+1}} \left( \sum_{k=T_j^1}^{T_j^z} e^{-\beta^{(i,j)} (v-T_j^k ) } \right) dv \notag \\
    & =\sum_{i=0}^{D-1} \sum_{j=0}^{D-1} \alpha^{(i,j)}  \sum_{z=1}^{\text{last}_j-1} \sum_{k=T_j^1}^{T_j^z} \left( e^{-\beta^{(i,j)} (T_j^{z+1}-T_j^k ) } - e^{-\beta^{(i,j)} (T_j^z-T_j^k ) } \right)  \notag\\
    &\quad + \sum_{i=0}^{D-1} \sum_{j=0}^{D-1} \alpha^{(i,j)} \sum_{k=T_j^1}^{T_j^z} \left( e^{-\beta^{(i,j)} (T_{\text{span}}-T_j^k ) } - e^{-\beta^{(i,j)} (T_j^{\text{last}}-T_j^k ) } \right) \notag \\
    &= - \sum_{i=0}^{D-1} \sum_{j=0}^{D-1} \frac{\alpha^{(i,j)}}{\beta^{(i,j)}}  \sum_{z=1}^{\text{last}_j} \left( e^{-\beta^{(i,j)} (T_{\text{span}}-T_j^z ) } - 1 \right) 
\end{align}

Therefore, we obtain the final form of \(\Gamma\) as follows:
\begin{equation}
    \Gamma=T_{\text{span}} \sum_{i=0}^{D-1} \theta_i - \sum_{i=0}^{D-1} \sum_{j=0}^{D-1} \frac{\alpha^{(i,j)}}{\beta^{(i,j)}}  \sum_{z=1}^{\text{last}_j} \left( e^{-\beta^{(i,j)} (T_{\text{span}}-T_j^z ) } - 1 \right) 
\end{equation}

\subsection{Simplification Algorithms for MDHP-GDS}\label{sec:appendix-simple-mdhp-gds}

\begin{algorithm}
\caption{Calculation of $\ln{L}: \text{Part1}$}
\begin{algorithmic}\label{algo:mdhp-gds-persudocode-lnl-part1}
    \REQUIRE{$\text{tMpT}, \alpha, \beta, \theta$}
    \ENSURE{$\text{Result}$}
    \STATE $\alpha\gets\text{expand}(\alpha,[:,1,:,1])$
    \STATE $\beta\gets\text{expand}(\beta,[:,1,:,1])$
    \STATE $\theta\gets\text{expand}(\beta,[:,1,:,1])$
    \STATE $A\text{exp}\gets\alpha*\exp{(\text{clamp\_max}(-\beta*\text{tMpT}, 80))}$
    \STATE $S\gets\text{sum}(A\text{exp}, \text{dim}=[2,3])$ 
    \STATE $\text{Result}\gets\text{sum}(\log{(\theta+S)})$ 
\end{algorithmic}
\end{algorithm}

\begin{algorithm}
\caption{Calculation of $\ln{L}: \text{Part2}$}
\begin{algorithmic}\label{algo:mdhp-gds-persudocode-lnl-part2}
    \REQUIRE{$T_\text{span}, \theta$}
    \ENSURE{$\text{Result}$}
    \STATE $\text{Result}\gets-T_\text{span}*\text{sum}(\theta)$
\end{algorithmic}
\end{algorithm}

\begin{algorithm}
\caption{Calculation of $\ln{L}: \text{Part3}$}
\begin{algorithmic}\label{algo:mdhp-gds-persudocode-lnl-part3}
    \REQUIRE{$T_\text{padded}, T_\text{span}, \alpha, \beta$}
    \ENSURE{$\text{Result}$}
    \STATE $T_\text{bcasted}\gets\text{expand}(T_\text{padded}, [:,dim,:])$
    \STATE $\beta\gets\text{expand}(\beta, [:,:,1])$
    \STATE $K_\text{inner}\gets\text{clamp\_max}(-\beta*(T_\text{span}-T_\text{bcasted}), 80)$
    \STATE $K_\text{inner}\gets\text{transpose}(\exp{(K_\text{inner})}, (0,1))$
    \STATE $K\text{sum}\gets\text{sum}(K_\text{inner}, \text{dim}=2)$
    \STATE $\text{Result}\gets\text{sum}(\alpha/\beta*K\text{sum})$
\end{algorithmic}
\end{algorithm}

\subsection{IP control function}\label{appendices_IP_control}
\textbf{PLA — Power-law acceleration strategy}: In \( f(t) \), \( a \) controls the  of IP growth, and \( b \) represents the growth rate over time.\\
\textbf{DEA — Delayed escalation attack strategy}: \( n_{\text{base}} \) represents the initial number of controlled IPs. \( k_1 \) indicates the linear growth rate in the early stage, and \( k_2 \) specifies the magnitude of the exponential increase in the later stage. \( t_1 \) again denotes the switching point, while \( \mu \) reflects the strength of the exponential growth.\\
\textbf{ASA — Adaptive stealth-oriented attack strategy}: \( n_{\text{max}} \) defines the upper bound of the number of controlled IPs. \( \gamma \) governs the IP growth rate, and \( t_0 \) is the midpoint of the sigmoid growth curve.\\
\textbf{DAM — Dynamic adjustment via multi-strategy integration}: \( n_{\text{base}} \) is the initial number of controlled IPs, and \( n_{\text{max}} \) is the maximum allowed number.  \( \alpha \in [0, 1] \) adjusts the weighting between two distinct growth modes. Here, \( \beta > 0 \) controls the rate of the first (more aggressive and explosive) growth phase. And \( \gamma > 0 \) governs the second phase, characterized by a smoother increase in the number of controlled IPs.

\begin{table}[!htbp]
\centering
\caption{IP control function (corresponds to the rate functions in Table \ref{tab:Attack strategy} on a row-by-row basis)} 
\setlength{\tabcolsep}{3pt}
\renewcommand{\arraystretch}{1.5}  
\begin{tabular}{c}
\hline
\textbf{IP control function} \\ \hline
$f(t)=\left(\frac{a}{b+1}\right) \cdot t^{b+1}$  \\ \hline
$
f(t)=
\begin{cases}
n_{base} + k_1\cdot t, & t < t_1, \\
n_{base} + k_1\cdot t_1 + k_2\cdot (e^{\mu(t-t_1)}-1), & t \ge t_1.
\end{cases}
$ \\ \hline
$f(t)= \frac{n_{max}}{1+e^{-\gamma(t-t_0)}}$ \\ \hline
$f(t)=n_{base} + (n_{max} - n_{base}) \cdot (\alpha e^{\beta}+(1-\alpha)(1-e^{-\gamma t}))$ \\ \hline
\end{tabular}
\label{tab: IP control function}
\end{table}

\subsection{MDHP-GDS Estimation Results Based on CDF Analysis}\label{Appendices_CDF}
\begin{figure}[htbp]
    \centering
\includegraphics[width=0.50\textwidth]{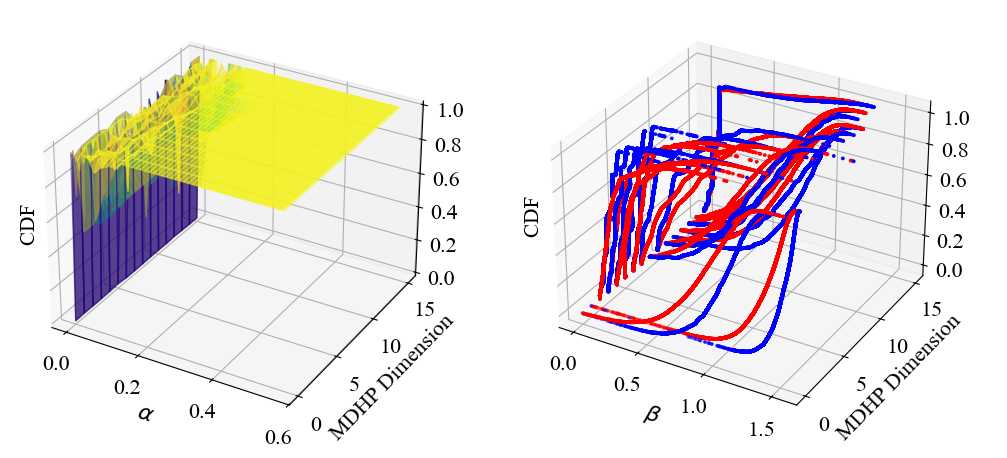}
    \caption{3D Visualization of CDF Dim 5.}
    \label{fig:experiment-cdf-5-3d}
\end{figure}

Figure \ref{fig:experiment-cdf-5-3d} presents three-dimensional CDF visualizations for $\alpha$ and $\beta$, focusing on interactions between all ECUs and ECU 5. The 3D representations reveal distinct distributional patterns between normal and attack states. The left subplot displays the excitation matrix ($\alpha$) CDF distribution, while the right subplot shows the decay matrix ($\beta$) CDF distribution, both exhibiting clear topological differences between attack and normal states. 

\begin{figure}[H]
    \centering
    \includegraphics[width=0.45\textwidth]{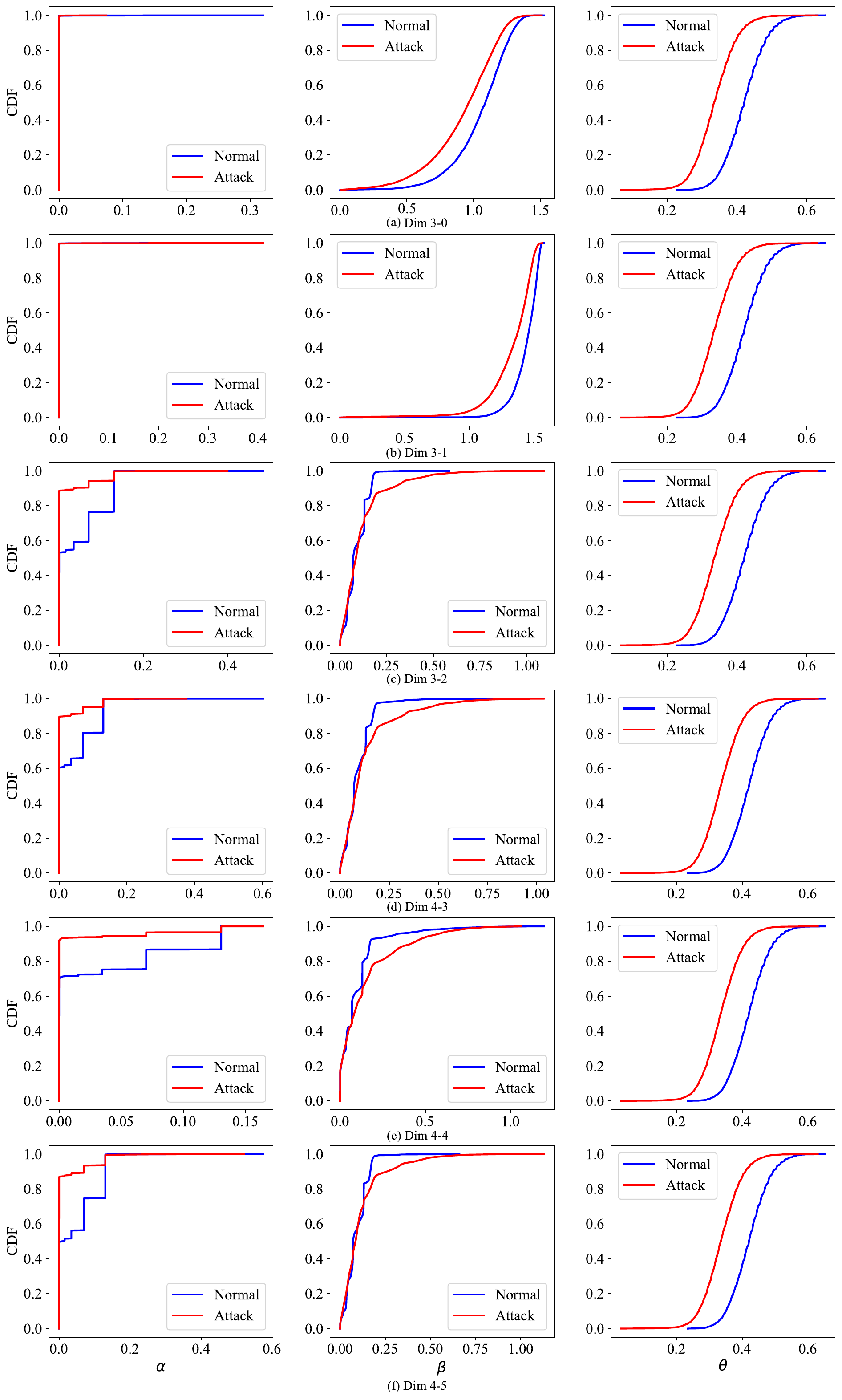}
    \caption{CDF of Different MDHP dimensions}
    \label{fig:appendix-cdf-2ds-many-dim}
\end{figure}

Figure \ref{fig:appendix-cdf-2ds-many-dim} shows the CDF curves in more MDHP dimensions. It is obvious that, in most cases, MDHP parameters can be a significant feature in classifying normal and attack message windows.

\subsection{Sampling algorithm}\label{algo:appendices_sampling_algorithm}
Detailed description of the sampling algorithm:\\
\textbf{NPP --- Nonhomogeneous Poisson Process}: This
algorithm implements a sampling procedure based on an nonhomogeneous Poisson process. To begin, a constant rate is used to sample candidate time points over the given interval(i.e., $\Delta t \gets Exponential(1/512)$)., ensuring that the temporal distribution of attack events remains consistent with realistic scenarios. For each candidate time point, an acceptance-rejection mechanism is applied: a sample is accepted with probability \( g(t)/g_{\text{max}} \). The final output consists of all accepted time points, which collectively follow the desired distribution.\\
\textbf{DRP --- Double random Poisson process}: This algorithm introduces a two-level randomness. First, we randomly generate an attack rate function$g(t)$, which originates exclusively from the Dynamic Adjustment via Multi-Strategy Integration strategy and is defined as:
$g(t) = w \cdot \alpha_1\,t^{\alpha_1 - 1} + (1 - w) \cdot \alpha_2\, e^{\alpha_2t}$ .
Here, the randomness stems from the random selection of the weight parameter $w$. To capture this first layer of randomness, the entire time horizon is divided into multiple intervals, and a distinct value of $w$ is randomly assigned to each interval. This results in a piecewise-varying rate function $g(t)$. Once the rate function is fixed, the thinning method is applied to sample the actual event times.\\
\textbf{ND --- Normal distribution}: This sampling process in this method is similar to that of the nonhomogeneous Poisson Process. The key difference lies in the selection of candidate time points, which are drawn from a normal distribution(i.e., $\Delta t \gets |\text{Normal}(\mu, \sigma)|$). The mean and variance of the distribution are predefined, ensuring that the timing of attacks remains consistent with realistic scenarios.

\begin{algorithm}
\caption{Sampling method based on nonhomogeneous Poisson Process}
\begin{algorithmic}\label{algo_sample_poisson}
\REQUIRE Time range $(t_{\min}, t_{\max})$, rate function $g(t)$, max rate $g_{\max}$
\ENSURE {Injection times array $\mathcal{A}$}

\STATE $t \gets t_{\min}$, $\mathcal{A} \gets [\,]$

\WHILE{$t < t_{\max}$}
    \STATE $\Delta t \gets \text{Exponential}(1 / 512)$
    \STATE $t \gets t + \Delta t$
    \IF{$t > t_{\max}$}
        \STATE \textbf{break}
    \ENDIF
    \STATE $u \gets \text{Uniform}(0, g_{\max})$
    \IF{$u < g(t)$}
        \STATE Append $t$ to $\mathcal{A}$
    \ENDIF
\ENDWHILE
\RETURN $\mathcal{A}$
\end{algorithmic}
\end{algorithm}

\begin{algorithm}
\caption{Sampling method based on Normal distribution}
\begin{algorithmic}\label{algo_sample_random_size}
\REQUIRE {Time range $(t_{\min}, t_{\max})$, rate function $g(t)$,  maximum rate $g_{\max}$}
\ENSURE {Injection times array $\mathcal{A}$}

\STATE $\mathcal{A} \gets [\,]$, $t \gets t_{\min}$

\STATE $\mu = 1 / 512 $, $\sigma = \mu / 2$

\WHILE{$t < t_{\max}$}
    \STATE $\Delta t \gets |\text{Normal}(\mu, \sigma)|$
    \STATE $t \gets t + \Delta t$
    \IF{$t > t_{\max}$}
        \STATE \textbf{break}
    \ENDIF
    \STATE $u \gets \text{Uniform}(0, 1)$
    \IF{$u < \dfrac{g(t)}{g_{\max}}$}
        \STATE Append $t$ to $\mathcal{A}$
    \ENDIF
\ENDWHILE

\RETURN $\mathcal{A}$
\end{algorithmic}
\end{algorithm}

\begin{algorithm}
\caption{Sampling method based on Double random Poisson process}
\begin{algorithmic}\label{algo_generate_random_intensity}
\REQUIRE Time range $(t_{\min}, t_{\max})$
\ENSURE Injection times array $\mathcal{A}$

\STATE \textbf{Function} $\text{generate\_random\_intensity}()$
\STATE \quad $g_{\text{mixed}}(t, w, \alpha_1, \alpha_2) \gets w \cdot \alpha_1 \cdot t^{\alpha_1 - 1} + (1 - w) \cdot \alpha_2 \cdot \exp(\alpha_2 \cdot t)$
\STATE \quad $intensity(t, break\_points, w, \alpha_1, \alpha_2) \gets$ 
\STATE \quad \quad \textbf{if} $t < break\_points[0]$ \textbf{then} 
\STATE \quad \quad \quad $g_{\text{mixed}}(t, w[0], \alpha_1, \alpha_2)$
\STATE \quad \quad \textbf{else if} $t < break\_points[1]$ \textbf{then} 
\STATE \quad \quad \quad $g_{\text{mixed}}(t, w[1], \alpha_1, \alpha_2)$
\STATE \quad \quad \textbf{else}
\STATE \quad \quad \quad $g_{\text{mixed}}(t, w[2], \alpha_1, \alpha_2)$
\STATE \quad $\alpha_1, \alpha_2 \gets 3.0, 4.0$ 
\STATE \quad $w \gets [\text{Uniform}(0, 0.33), \text{Uniform}(0.33, 0.66),\text{Uniform}(0.66, 1)]$
\STATE \quad $intensity\_func \gets \text{partial}(intensity, break\_points, w, \alpha_1, \alpha_2)$
\STATE \quad $g\_max \gets \text{max}(\text{evaluate} \, g_{\text{mixed}}(t))$

\STATE \textbf{Return} $intensity\_func, g\_max$

\STATE \textbf{Function} $\text{sample}(time\_range)$
\STATE \quad $g, g_{\text{max}} \gets \text{generate\_random\_intensity}()$, $t \gets start$, $\mathcal{A} \gets [\,]$
\WHILE{$t < end$}
    \STATE $\Delta t \gets \text{Exponential}(1 / 512)$
    \STATE $t \gets t + \Delta t$
    \IF{$t > end$} 
    \STATE \textbf{break}
    \ENDIF
    \STATE $u \gets \text{Uniform}(0, g_{\max})$
    \IF{$u < g(t)$}
        \STATE Append $t$ to $\mathcal{A}$
    \ENDIF
\ENDWHILE
\RETURN $\mathcal{A}$
\end{algorithmic}
\end{algorithm}

\end{document}